\documentclass[sigconf]{acmart}

\def\BibTeX{{\rm B\kern-.05em{\sc i\kern-.025em b}\kern-.08emT\kern-.1667em\lower.7ex\hbox{E}\kern-.125emX}}
    
\copyrightyear{2019} 
\acmYear{2019} 
\setcopyright{acmlicensed}
\acmConference[ARES '19]{Proceedings of the 14th International Conference on Availability, Reliability and Security (ARES 2019)}{August 26--29, 2019}{Canterbury, United Kingdom}
\acmBooktitle{Proceedings of the 14th International Conference on Availability, Reliability and Security (ARES '19), August 26--29, 2019, Canterbury, United Kingdom}
\acmPrice{15.00}
\acmDOI{10.1145/3339252.3340500}
\acmISBN{978-1-4503-7164-3/19/08}

\usepackage{todonotes}
\usepackage{threeparttable}
\usepackage{graphics}
\usepackage{booktabs}
\usepackage{multirow}
\usepackage{makecell}
\usepackage{amsmath}
\usepackage{adjustbox}
\usepackage{amssymb}
\usepackage{pifont}
\usepackage[ruled,vlined]{algorithm2e}

\begin{document}

%
\title[Privacy-Enhancing Fall Detection from Remote Sensor Data\\Using Multi-Party Computation]{Privacy-Enhancing Fall Detection from Remote Sensor Data Using Multi-Party Computation}

%
\author{Pradip Mainali}
\authornote{Both authors contributed equally to this research.}
\email{pradip.mainali@onespan.com}
\affiliation{%
  \institution{OneSpan}
  \city{Brussels}
  \state{Belgium}
  \postcode{1853 Grimbergen}
}

\author{Carlton Shepherd}
\authornotemark[1]
\email{carlton.shepherd@onespan.com}
\affiliation{%
  \institution{OneSpan}
  \city{Cambridge}
  \state{United Kingdom}
}

\begin{abstract}
Motion-based fall detection systems are concerned with detecting falls from vulnerable users, which is typically performed by classifying measurements from a body-worn inertial measurement unit (IMU) using machine learning. Such systems, however, necessitate the collection of high-resolution measurements that may violate users' privacy, such as revealing their gait, activities of daily living (ADLs), and relative position using dead reckoning. In this paper, we investigate the application of multi-party computation (MPC) to IMU-based fall detection for protecting device measurement confidentiality. Our system is evaluated in a cloud-based setting that precludes parties from learning the underlying data using multiple, disparate cloud instances deployed in three geographical configurations. Using a publicly-available dataset, we demonstrate that MPC-based fall detection from IMU measurements is practical while achieving state-of-the-art error rates. In the best case, our system executes in 365.2 milliseconds, which falls well within the required time window for on-device data acquisition (750ms).
 
\end{abstract}

%
%

\ccsdesc[500]{Security and privacy~Security services}
\ccsdesc[300]{Security and privacy~Distributed systems security}
\ccsdesc[300]{Security and privacy~Domain-specific security and privacy architectures}
\ccsdesc[100]{Computer systems organization~Cloud computing}


%
\keywords{Fall detection, multi-party computation (MPC), Internet of Things (IoT), cloud computing}

\maketitle

\section{Introduction}
\label{sec:intro}

Falls are the leading cause of fatal injury and the most common cause of nonfatal, trauma-related hospital admissions among older adults, with 20--30\% of mild-to-severe injuries and 40\% of injury-related fatalities~\cite{musci2018online}.  The annual medical costs attributable to fatal and nonfatal falls was estimated at \$50bn (USD) in 2015~\cite{florence2018medical}.  Fall detection systems are concerned with automatically detecting the occurrence of falls within the homes of elderly and disabled users, with the aim of minimising response times to potential injuries and facilitating independent living~\cite{mubashir2013survey}.  These systems involve the collection of data, such as accelerometer data from a body-worn device~\cite{mauldin2018smartfall,liu2012fall,santoyo2018analysis,doukas2007patient,yavuz2010smartphone,albert2012fall,luvstrek2009fall}, video imagery from cameras in the home~\cite{zhang2012privacy,mastorakis2014fall,kwolek2014human,edgcomb2012automated}, and proximity sensors in ceilings~\cite{tao2012privacy}. This data is typically classified with respect to a dataset of known fall and non-fall data using machine learning.
 
The deployment challenges surrounding fall detection systems have, to date, remained largely out-of-scope in current work, which implicitly assume a local deployment or focus solely on classification performance using off-line analyses. Recent work has touted the benefits of a cloud-based, software-as-a-service (SaaS) approach for mobility-based sensing systems, where applications and their underlying platform can be updated, upgraded, monitored and secured with greater flexibility~\cite{doukas2011managing,fortino2014bodycloud,gravina2017cloud}. This is important in managing large numbers of remote devices, where patch penetration remains a significant challenge~\cite{palani2016invisible}. Indeed, failing to update devices could lead to potentially unsafe misclassification errors due to accuracy discrepancies between implementations.  

This principle of \emph{algorithm agility} provides greater flexibility in changing the underlying classification algorithm in a cloud-based scenario without potential penetration issues of patching and updating large numbers of devices in the field.  Another benefit lies in the separation of the platform development, e.g. device hardware, firmware and software, from the algorithm implementation itself. In this paradigm, individual device OEMs\footnote{OEM: Original Equipment Manufacturer.} are relinquished from the responsibility of implementing and maintaining the detection algorithm. Rather, this can be performed by a dedicated third-party with the potential of supporting devices from multiple OEMs in a service-oriented architecture, which we focus upon in this work.
 

 
A myriad of security and privacy concerns remain, however, regarding large volumes of data being harvested, used or, indeed, misused by SaaS providers~\cite{zhang2014iot}. At worst, sensor measurements collected from users' devices could be exploited to infer their activity patterns and whereabouts, with major privacy implications.  This is compounded by the risks of disclosing plaintext measurements, say, after the exploitation of a vulnerable service, such as insecure AWS S3 buckets~\cite{continella2018there}. A large corpus of work has already demonstrated identifying users based on their gait from accelerometer and other inertial measurement unit (IMU) data~\cite{gafurov2007gait,thang2012gait,mantyjarvi2005identifying,juefei2012gait,gafurov2007survey}; their position using dead reckoning~\cite{pratama2012smartphone,kang2015smartpdr,wang2018pedestrian}; and determining activities of daily living (ADLs), like whether the user is walking, sleeping, and sitting~\cite{stikic2008adl,hong2010mobile}.   Existing work on privacy-enhancing fall detection has focussed on non-cryptographic techniques for video-based methods, including image blurring, silhouetting and foreground extraction. However, these still reveal significant information about users, such as their presence. Moreover, they do not generalise to the significant proportion of fall detection proposals using time-series IMU measurements from personal devices, such as smartphones and  smartwatches~\cite{mauldin2018smartfall,santoyo2018analysis,cao2012falld,vilarinho2015combined,albert2012fall}.

A promising solution is multi-party computation (MPC), where measurements are split using a linear secret sharing scheme (LSSS) and distributed to mutually distrusting parties. After this, functions can be jointly computed without parties disclosing their plaintext inputs to others. Existing literature work has already shown that machine learning inferencing, upon which many fall detection systems rely, can be performed efficiently using MPC. This includes MPC-based image recognition using support vector machines (SVMs)~\cite{epic} and remote rehabilitation treatment classification from patient data using decision trees~\cite{de2014practical}.  

In this work, we present the implementation and evaluation of the first system for perform MPC-based fall detection from IMU device data in order to protect measurement confidentiality. MPC is used to perform the requisite feature extraction and classification without the parties learning the contents of IMU measurements received from the device. We evaluate the system, illustrated in Figure \ref{fig:high_level}, using an off-the-shelf device and three MPC parties deployed using a public cloud service.  In short, this work presents the following contributions:

\begin{figure}
    \centering
    \includegraphics[width=\linewidth]{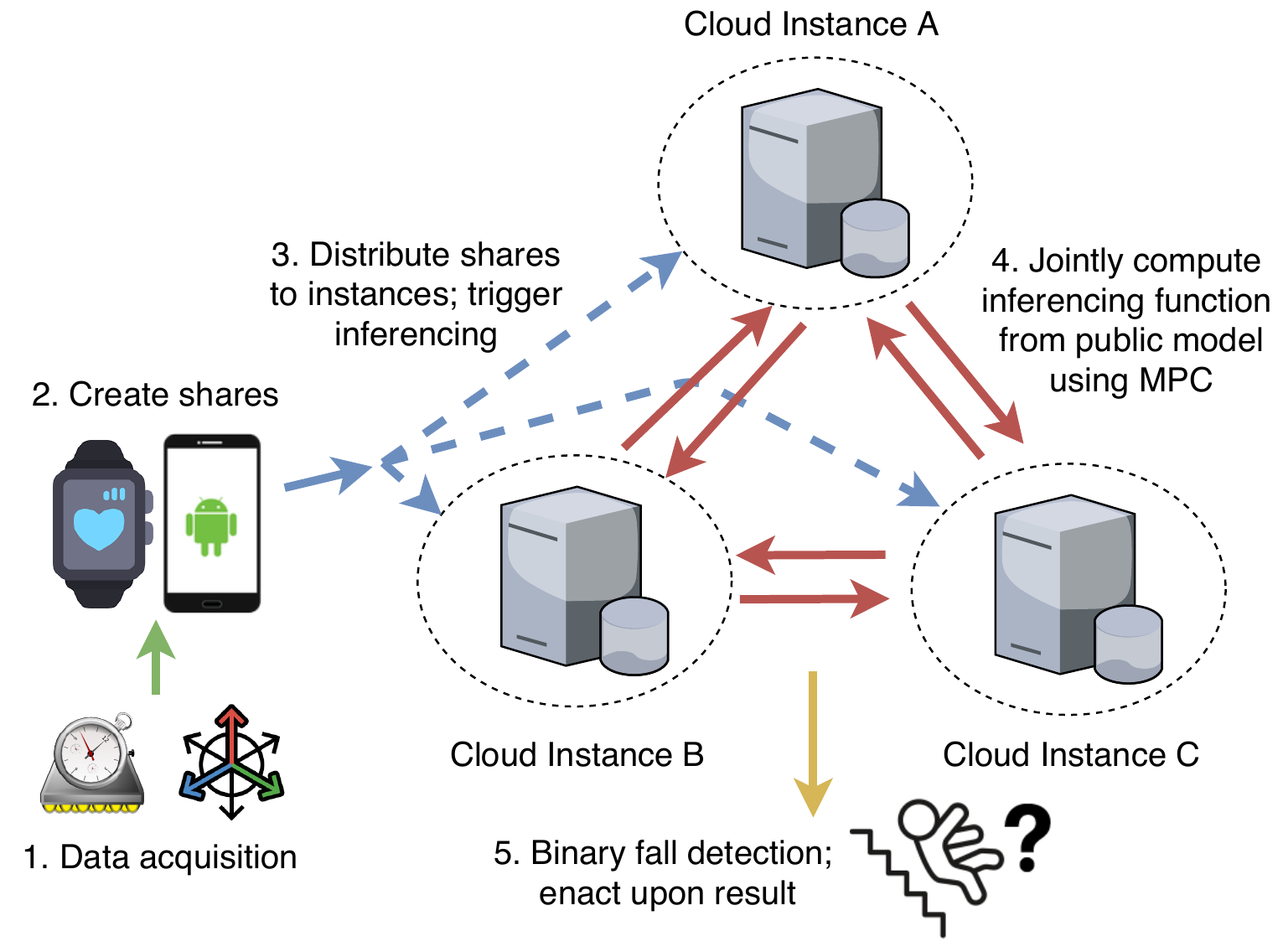}
    \caption{High-level workflow for MPC-based fall detection.}
    \label{fig:high_level}
    \vspace{-0.4cm}
\end{figure}


\begin{itemize}
    \item The first framework for performing privacy-enhancing fall detection from device IMU measurements using MPC. The proposal preserves measurement confidentiality in a cloud-based model while performing feature extraction and classification using machine learning.
    \item A two-part evaluation with experiments conducted using a publicly-available fall detection dataset. Our inferencing framework executes faster than the time necessary for on-device measurement acquisition, and achieves error rates commensurate with state-of-the-art schemes.
\end{itemize}

\subsection{Document Structure}

This paper proceeds with a general review of fall detection systems in Section \ref{sec:fall_detection}, which is followed by a discussion of the privacy challenges associated with existing systems. In Section \ref{sec:private_falldetection}, we present the assumptions and threat model in greater detail, before introducing and motivating the use of MPC as a solution to the aforementioned challenges.  This section also develops the design of the proposed framework after which, in Section \ref{sec:imp}, the implementation details are discussed.  Section \ref{sec:eval} presents an evaluation of the proposed framework, comprising a discussion of the methodology and experimental results in the form of classification and computational performance.  Lastly, we conclude our work in Section \ref{sec:conc}, including a discussion of future research directions.

\section{Fall Detection}
\label{sec:fall_detection}

Fall detection systems fall broadly into two categories: 1), those using devices situated in the user's ambient environment; and, 2), those using body-worn devices (BWDs), e.g. smartwatches.  We now describe prominent existing schemes and their operation. This section is not an exhaustive analysis---for this, the reader is referred to Mubashir et al.~\cite{mubashir2013survey} and Delahoz et al.~\cite{delahoz2014survey}---but, rather, we aim to summarise the salient approaches and results.

\subsection{Ambient Device-based Systems}
Privacy-enhancing fall detection has focused almost entirely on non-cryptographic techniques from ambient-based devices, such as silhouetting, blurring and foreground masking from camera data. Mastorakis and Makris~\cite{mastorakis2014fall} and Zhang et al.~\cite{zhang2012privacy} employ RGBD data from a Microsoft Kinect to detect five ADLs, using only the depth values and foreground mask of the RGB data, thus occluding users' facial features.  Tao et al.~\cite{tao2012privacy} study the use of a ceiling-based sensor network comprising infrared cameras for fall detection in a home environment.  The sensors output binary values to indicate the existence of persons underneath, which are used to model a map of the user's home.  Edgcomb and Vahid~\cite{edgcomb2012automated} explore four techniques for privacy-enhancing video-based fall detection: \emph{blurring} (of the user), \emph{silhouetting}, and covering the user with a graphical \emph{box} and \emph{oval}. Raw video imagery is first taken from a stationary in-home camera, before applying one of these techniques.

\subsection{BWD-based Systems}
The use of ambient-based devices has inherent issues associated with detecting falls in occluded areas: fitting multiple cameras to cover all possible locations in the user's living environment---bathroom, bedroom, kitchen, and so on---may impose significant costs relating to unit installation and maintenance.  Another paradigm employs devices likely to be already owned and carried by the user, such as a smartwatch or smartphone.  Unlike the previous section, we are unaware of proposals that investigate privacy-enhancing fall detection from BWDs to the best of our knowledge.  As such, we provide a general review of highly-cited and state-of-the-art work from the literature.

Mauldin et al.~\cite{mauldin2018smartfall} present SmartFall, which uses accelerometer data sampled at 31.25Hz from a user-worn smartwatch to detect user falls.  Smartwatch data is streamed to the user's smartphone that extracts features pertaining to the vector magnitude of the accelerometer data and the difference between the maximum and minimum magnitudes over a 750ms sliding window (Equations \ref{eq:mag} and \ref{eq:delta_s}). Na\"{i}ve Bayes, Support Vector Machine (SVM) and Recurrent Neural Network (RNN) classifiers are evaluated using labelled data from seven volunteers between 21--55 years old, alongside the Farseeing dataset~\cite{mellone2012smartphone} with data from 2,000 participants.  Results of 0.37--0.79 precision, 0.55--1.0 recall, and 0.65--0.99 accuracy are reported, depending on the chosen dataset and classifier.

\begin{equation}
    |\vec{v}| = \sqrt{x^2 + y^2 + z^2}
    \label{eq:mag}
\end{equation}

\begin{equation}
    \Delta s = \max(|\vec{v_i}|) - \min(|\vec{v_j}|), i \neq j
    \label{eq:delta_s}
\end{equation}

Liu and Cheng~\cite{liu2012fall} investigate IMUs placed at users' waists that collects accelerometer measurements at 200Hz. The vector magnitude is computed, along with the fast changed vector, vertical acceleration and posture angle between the vertical acceleration and gravity over a 0.1s sliding window.  The features are inputted to an SVM with labelled fall data from 15 volunteers who performed 10 simulated falls and 11 ADLs repeated 10 times. The proposal yields error rates of 98.38\% accuracy, 97.40\% recall, and 99.27\% precision.

Santoyo-Ram{\'o}n et al.~\cite{santoyo2018analysis} investigate IMUs attached to the user's chest, waist, wrist and thigh---the data from which is aggregated into 15 second segments by a smartphone. A sliding window of 0.5s is used in which the fast changed vector of accelerometer measurements is computed alongside the magnitudes' mean, standard deviation and mean absolute distance, and the mean rotation angle and mean module of the $(Y, Z)$ and $(X, Z)$ components.  The features are inputted to SVM, k-Nearest Neighbour, Na\"{i}ve Bayes and Decision Tree classifiers, and evaluated using data from 19 subjects who performed 746 ADLs and falls. The geometric mean of the precision and recall is used as the performance metric, with results of 0.614--0.999 depending on the algorithm and sensor combination.

Doukas et al.~\cite{doukas2007patient} investigate the use of an accelerometer attached to the user's foot for fall detection.  The sensor data is transmitted to a receiving device, such as a laptop, where it is normalised and inputted to an SVM classifier in raw form after applying a Kalman filter.  The evaluation reports a classification accuracy of 98.2\% using a limited set of labelled data from two volunteers. 

Cao et al.~\cite{cao2012falld} present E-FallD, which uses accelerometer data from an Android smartphone. The vector magnitude of the accelerometer components is computed and a global threshold is fixed for identifying falls based on the phone's measured acceleration.  The work also investigates multiple thresholds customised to the user's gender, age, and body mass index (BMI). The authors recruit 20 participants of varying gender, age and BMI, who perform a total of 400 falls and 1200 ADLs. The approach yields results of 86.75\% precision and 85.5\% recall using global thresholds, and 92.75\% precision and 86.75\% recall with customised thresholds.

Similarly, Vilarinho et al.~\cite{vilarinho2015combined} use accelerometer data from an Android smartphone and smartwatch.  Here, a rule-based system is devised for threshold-based detection using the vector magnitude, fall index, and absolute vertical acceleration as features over a 0.8s sliding window. The approach is evaluated using data from three participants, who performed 12 falls and 7 ADLs each, with a reported 0.68 classification accuracy, 0.78 precision, and 0.63 recall.

Yavuz et al.~\cite{yavuz2010smartphone} propose another system based on accelerometer values from an Android smartphone. In contrast, the authors use solely frequency domain features, namely the wavelet coefficients after applying a discrete wavelet transform. A threshold-based approach is used to test the coefficients generated from the phone against those of previously observed falls.  100 fall sequences from five volunteers were collected to evaluate the system, with results of 46-95\% precision and 88-90\% recall.

Albert et al.~\cite{albert2012fall} use accelerometer values from an LG G1 Android smartphone located at the user's pelvic region. 178 features across several categories are subsequently extracted: moments, e.g. kurtosis and skew; moments between successive samples; smoothed root mean squares; extremities, e.g. min and max; histogram values; Fourier components; mean acceleration magnitude; and cross products.  The features are inputted to five classifiers---SVM, Logistic Regression, Na\"{i}ve Bayes, Decision Tree and kNN---and evaluated using data from 18 simulated falls from 15 subjects, resulting in 98\% classification accuracy.

Musci et al.~\cite{musci2018online} evaluate the use of recurrent neural networks (RNNs) and long short-term memory units (LSTMs), and propose a deep learning architecture for ternary fall detection for whether or not a fall has occurred, and whether one is about to occur.   The proposal is evaluated using the SisFall dataset from \cite{sucerquia2018real}, comprising accelerometer and gyroscope measurements from a smartphone and a custom IMU attached to the waists of 38 users; the dataset is labelled with respect to the three aforementioned states.  A confusion matrix of the three classes reports 92.86--97.16\% classification accuracy following hyperparameter optimisation.

Lu{\v{s}}trek and Kalu{\v{z}}a~\cite{luvstrek2009fall} explore the use of a body area network (BAN) comprising 12 IMUs at the user's shoulders, wrists, hips, knees and ankles.  Features are collected regarding the coordinates of units in a body coordinate system, their velocities, absolute distance, and relative angles.  The features are used as input to seven machine learning algorithms---SVM, DT, kNN, NB, Random Forest, Adaboost, and bagging---implemented in the Weka library. Three participants were recruited who performed a total of 45 activities: 15 falls and 10 lying down, sitting down and walking ADLs each.  73.4--96.3\% classification accuracy is reported depending on the algorithm used, with SVM yielding the best case.

\subsection{Discussion}

Generally, fall detection systems rely upon traditional supervised learning algorithms and, more recently, deep neural networks (LSTMs and RNNs), trained over labelled fall data collected \emph{a priori}.  The majority of existing work models binary classification, i.e. `fall' or `not fall', while other work also models whether a fall \emph{may} be about to occur and classification of general ADLs, such as sitting, walking, and lying down. 


\begin{figure}
    \centering
    \includegraphics[width=0.85\linewidth]{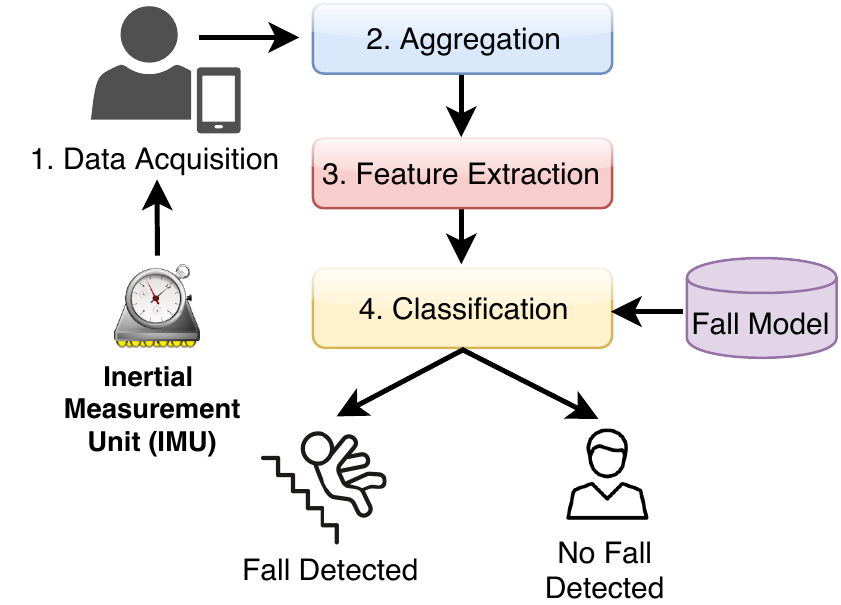}
    \caption{Overview of IMU-based binary fall detection.}
    \label{fig:fall_detection}
    \vspace{-0.5cm}
\end{figure}

Additionally, the generation of \emph{global} models that generalise to all users, rather than individuals, is also a common facet of current work.  For BWD-based systems, accelerometer measurements are used ubiquitously from which a variety of features are extracted prior to classification, which is illustrated in Figure \ref{fig:fall_detection} for the binary case. Most work focuses on  time-domain features, such as the vector magnitude, arithmetic mean, standard deviation, and maximum and minimum values. The choice of algorithm remains varied, although SVMs tend to exhibit the best performance with 0.9838~\cite{liu2012fall}, 0.982~\cite{doukas2007patient} and 0.73--0.99~\cite{mauldin2018smartfall} accuracy. Recent work \cite{mauldin2018smartfall,musci2018online} has begun to explore the application of deep learning using LSTMs and RNNs with promising results---0.9286--0.9716~\cite{musci2018online} and 0.85--0.99~\cite{mauldin2018smartfall} accuracy---but work in the area remains limited.

Rather than using machine learning for the discovery of optimal decision boundaries, another widely-used approach has been the use of manually set threshold values, which are discovered experimentally. Proponents of this approach have touted faster classification as an advantage over machine learning-based systems, and thus more suitable for constrained devices; however, such approaches generally yield higher error rates and are evaluated over limited datasets, e.g. with five~\cite{yavuz2010smartphone} and three~\cite{vilarinho2015combined} participants.  We present a summary of BWD-based systems in Table \ref{tab:related}.


\begin{table}
\centering
\caption{Comparison of BWD-based fall detection systems.}
\resizebox{\columnwidth}{!}{%
\begin{threeparttable}
\renewcommand{\arraystretch}{1.3}

\begin{tabular}{@{}ccc@{}}

\toprule
\textbf{Proposals} & \textbf{Algorithms} & \textbf{Error Rates} \\ \midrule
SmartFall~\cite{mauldin2018smartfall} & NB, SVM, RNN & \makecell{0.37--0.79 (P), 0.55--1.0 (R),\\0.65--0.99 (A)} \\[0.4cm]
Liu \& Cheng~\cite{liu2012fall} & SVM & \makecell{0.9927 (P), 0.9740 (R),\\0.9838 (A)} \\[0.5cm]
S-R et al.~\cite{santoyo2018analysis} & \makecell{SVM, kNN,\\NB, DT} & 0.614--0.999 (GMPR) \\[0.3cm]
Doukas et al.~\cite{doukas2007patient} & SVM & 0.982 (A) \\[0.3cm]
E-FallD~\cite{cao2012falld} & Thresholds & \makecell{{\it GT}: 0.8675 (P), 0.855 (R)\\{\it CT}: 0.9275 (P), 0.8675 (R)}\\[0.3cm]
Vilarinho et al.~\cite{vilarinho2015combined} & Thresholds & \makecell{0.78 (P), 0.63 (R), 0.68 (A)}\\[0.3cm]
Yavuz et al.~\cite{yavuz2010smartphone} & Thresholds & 0.88--0.90 (P), 0.46--0.95 (R) \\[0.3cm]
Albert et al.~\cite{albert2012fall} & \makecell{SVM, LR, NB,\\DT, kNN} & 0.982 (A) \\[0.3cm]

Musci et al.~\cite{musci2018online} & RNN, LSTM & 0.9286--0.9716 (A) \\[0.3cm]
Lu{\v{s}}trek \& Kalu{\v{z}}a~\cite{luvstrek2009fall} & \makecell{SVM, DT, kNN,\\NB, RF, Adaboost,\\Bagging} & 0.732--0.963 (A) \\ \bottomrule
\end{tabular}

\begin{tablenotes}
\item P: Precision, R: Recall, A: Classification accuracy, GMPR: Geometric mean of precision and recall.
\end{tablenotes}
\end{threeparttable}
}
\label{tab:related}
\end{table}

\section{Privacy-Enhancing BWD-Based Fall Detection}
\label{sec:private_falldetection}



To the best of our knowledge, privacy-enhancing BWD-based fall detection is yet to be investigated.  Accelerometers, used in the bulk of current proposals, measure fine-grained movements at high frequency rates (typically $>$30Hz) on commodity devices, such as smartphones and smartwatches. A wealth of literature exists surrounding the identification of users based on their gait using accelerometer measurements alone from such devices~\cite{gafurov2007gait,thang2012gait,mantyjarvi2005identifying,juefei2012gait,gafurov2007survey}. Furthermore, we are also aware of IMU measurements being used for recognising non-fall ADLs~\cite{stikic2008adl, hong2010mobile} and performing user position tracking using dead reckoning~\cite{pratama2012smartphone,kang2015smartpdr,wang2018pedestrian}, which could be exploited to violate user privacy.

\subsection{Applying Multi-Party Computation (MPC)}

We observe that MPC can be applied to acquire data, extract features, and perform fall classification in a way that preserves IMU measurement confidentiality.  In this work, we tackle the base case of MPC using arithmetic secret sharing from Shamir's secret sharing scheme (SSSS)~\cite{shamir1979share}, which is briefly described as follows.

\subsubsection{Shamir's Secret Sharing Scheme}

SSSS divides a secret $\mathcal{S}$ into $n$ shares, $\textbf{s} = \{s_1, s_2, \dots, s_n\}$, such that knowing $t$ shares from $\textbf{s}$ makes $\mathcal{S}$ easy to compute, but knowing fewer shares than $t$ renders $\mathcal{S}$ undetermined. The scheme relies on the fact that $t$ points are required to construct a polynomial of degree $t-1$, which is defined in the field $\mathbb{F}_P$.  Next, $t-1$ positive integers, $a_1,\dots,a_{t-1}$ are chosen with $a_i < P$, with the secret assigned as $a_0 = \mathcal{S}$. The polynomial $f(x) = a_0 + a_{1}x + a_{2}x^{2} + a_{3}x^{3} + \cdots + a_{t-1}x^{t-1}$ is built with which $n$ points are constructed from $i=1,\dots,n$, to retrieve pairs $(i, f(i))$.  Any subset with $t$ distinct pairs can be used to recover the constant term, $a_0$, i.e. the secret, which is found using Lagrange interpolation.  

\subsubsection{MPC from Arithmetic Secret Sharing}
\label{sec:mpc_fund}
An important observation is that SSSS allows each party to locally compute linear combinations of secrets and public values. Firstly, let us define $[x]_i$ as the $i^{th}$ share of data item, $x$. SSSS allows parties to locally compute the following arithmetic operations over their shares~\cite{chen:ping}:
\begin{itemize}
        \item \emph{Addition} ($[c] \gets [a] + [b]$): Each party $p_i$ can compute a new share, $[c]_i$, that is the addition of two other shares, i.e. $[c]_i = [a]_i + [b]_i$.
    \item \emph{Addition of a secret and public value}  ($[c] \gets [a] + \alpha, \alpha \in \mathbb{F}_P$): Party $p_i$ can compute a new share from the addition of a held share and a public value $\alpha \in \mathbb{F}_P$; that is, $[c]_i = [a]_i + \alpha$.
    \item \emph{Multiplication of a secret and public value} ($[c] \gets [a] \cdot \alpha, \alpha \in \mathbb{F}_P$): $p_i$ can compute a new share of the multiplication of a held share with a public value, $[c]_i = [a]_i \cdot \alpha$.
\end{itemize}

Computing a new share that is the multiplication of two others, $[c] = [a] \cdot [b]$, is less straightforward: given that SSSS uses polynomials of degree $t$, then $[a\cdot b] \neq [a]_i\cdot[b]_i$, as the resulting polynomial has degree $2t$. A method for computing $t$-degree polynomials representing $[c] \gets [a]\cdot[b]$ was presented in \cite{ben1988completeness}. As opposed to entirely local computation for addition of shares and the addition and multiplication of a share with a constant value, multiplying shares necessitates one round of communication.  While many functions can be computed using only these operations, they preclude integer comparison, e.g. $=$, $\neq$, $<$, and $>$, and Boolean operations, e.g. {\tt AND} and {\tt OR}.  The reader is referred to \cite{chen:ping} for a comprehensive description and analysis of these protocols; the differences in asymptotic communication complexity are reproduced from \cite{chen:ping} in Table \ref{tab:comm}. 


\begin{table}
\centering
\caption{Communication cost for MPC functions using arithmetic secret sharing~\cite{chen:ping}.}
\begin{threeparttable}
\begin{tabular}{cc}
\toprule
\textbf{Function} & \makecell{\textbf{Communication}\\\textbf{Cost} (in rounds)} \\ \midrule
Multiplication & 1 \\
Inner Product & 1 \\
Inverse & 1 \\
Random element & 1 \\
Random bit & 3 \\
Equality ($=$) & 2 \\
$\geq$ & $2m+4$\\
$\leq$ & $2m+4$\\
{\tt AND}, {\tt OR} & $2\log(k)$\\\bottomrule
\end{tabular}
\begin{tablenotes}
\item Where $k$ is the number of operands and $m$ is the bound $x,y \in [0,2^{m}-1]$ wherein operands $x$ and $y$ lie.
\end{tablenotes}
\end{threeparttable}
\label{tab:comm}
\end{table}



\subsection{Assumptions and Threat Model}
\label{sec:threat_model}

An outsourcing model is assumed whereby a body-worn device streams IMU measurements to a cloud-based service that performs the feature extraction and classification.  The focus is on binary classification for inferring the occurrence of falls using a globally-trained model in line with the bulk of existing work from Section~\ref{sec:fall_detection}. We divide the entities involved into two categories.

\subsubsection{Source device}

We assume the existence of a body-worn device---a smartphone in the user's pocket, smartwatch, a sensor placed on the user's belt, or otherwise---with an IMU. The device is also assumed to possess the ability to split measurements into shares using SSSS and stream them to an Internet-facing service over a secure channel, such as TLS using a pre-installed certificate. In this work, the IMU measurements are considered to be trusted; the focus is on the unauthorised disclosure of measurements after departing the device over a network to a cloud-based service. 

\subsubsection{Cloud service}

 Comprises the software components---database instance, REST APIs, and so on---for receiving, storing and classifying falls from IMU sensor data on a public cloud platform. The primary threat is the exfiltration and disclosure of raw IMU measurements from a remote attacker, such as from an unsecured, public-facing database, e.g. AWS S3 bucket, which could be used to model and infer users' general ADLs, position, or gait for user identification. This work tackles the base case for MPC where parties operate under an honest-but-curious model, whereby the protocol specification is executed as intended.  We also assume that the output, i.e. whether or not a fall occurred, is learned by the cloud services in order to contact an emergency number or provide alternative remediation. 
 We note that \emph{some} information may be revealed by the communication between the device and the cloud service, such as the absence of receiving protected IMU data at certain times of day, e.g. at night where the user deactivates their smartphone or smartwatch before sleeping. However, this is orthogonal to the ultimate goal of this work of preventing either the service or an external adversary from learning additional information from IMU measurements beyond that intended for detecting falls.


\subsection{Workflow}
\label{sec:workflow}

The principle stages of the proposed system were illustrated previously in Figure \ref{fig:high_level} and described as follows:

\begin{enumerate}
    \item \emph{Data acquisition}. The collection of tri-axial IMU data on the device, e.g. smartphone, at an appropriate sampling frequency. This may be unimodal input, i.e. \emph{only} accelerometer~\cite{liu2012fall,mauldin2018smartfall,cao2012falld,doukas2007patient,yavuz2010smartphone}; multi-modal, e.g. accelerometer and gyroscope together~\cite{santoyo2018analysis}; or aggregating measurements from multiple sensors in a body-area network~\cite{luvstrek2009fall}.
    \item \emph{Secret sharing}. Given a tri-axial IMU sample, $[A_x, A_y, A_z]$, the master device splits each component into $n$ shares using SSSS. In this work, we set $n$ equal to the number of authorised communicating parties, while the reconstruction threshold is fixed at $t=3$.  In short, IMU samples are split as follows:
    \[
\textbf{D}=
  \begin{pmatrix}
    [A_x]_0 & [A_x]_1 & [A_x]_2 & \cdots & [A_x]_n \\
    [A_y]_0 & [A_y]_1 & [A_y]_2 & \cdots & [A_y]_n \\
    [A_z]_0 & [A_z]_1 & [A_z]_2 & \cdots & [A_z]_n 
  \end{pmatrix}
\]
Where the rows comprise the individual shares of a particular component, while each column represents the $i^{th}$ share of all components.
    \item \emph{Share distribution}. The device individually transmits the column elements from $\textbf{D}$ to each cloud service instance; that is, $[A_x]_0$, $[A_y]_0$ and $[A_z]_0$, is sent to party $P_0$, $[A_x]_1$, $[A_y]_1$ and $[A_z]_1$ to $P_1$, and so on. By sharing each sample independently, no single cloud-based party may recover the underlying IMU samples assuming the reconstruction threshold above. 
    \item \emph{MPC-based inferencing}.  The parties use arithmetic operations upon the received shares to perform fall detection feature extraction, such as computing the vector magnitude of IMU samples, and subsequent classification using a desired machine learning classifiers, e.g. support vector machine (SVM), logistic regression (LR), and na\"{i}ve Bayes (NB).
    \item \emph{Classification decision}. The resulting label, $\textbf{L} \in \{0,1\}$, is used to denote the occurrence of a fall in a binary fashion. This value may be used to raise an alarm or telephone an emergency contact number.
\end{enumerate}





\section{Implementation}
\label{sec:imp}

This section describes the implementation of the device- and cloud-side systems introduced in the previous section.

\subsection{Device Application}
\label{sec:device_app}

A Python application was developed that reads CSV IMU data line-by-line from file and creates shares of each sensor modality component using SSSS.  The resulting shares are base 64-encoded and JSON-formatted, and transmitted over TLS to their respective cloud parties using the method described in Section \ref{sec:workflow}. The application was deployed on a Raspberry Pi 3 Model B, with a Broadcom BCM2837 system-on-chip with a quad-core ARM Cortex-A53 at 1.2GHz, 1GB RAM, and Broadcom BCM43438 WiFi module (802.11N, 150 Mbit/s at 2.4GHz).  The board was connected to a corporate WiFi network over which all communications occurred. Benchmarking procedures were also implemented using the Python \texttt{time} module, which wraps functions from the GNU C standard library, for measuring the upload time for transmitting the shares and receiving an acknowledgement from the server.

\subsection{Cloud Framework}
\label{sec:cloud_imp}

We developed a cloud-based framework for performing MPC using arithmetic secret sharing comprising three applications.

    \textbf{\emph{Web server}}. A Python web server application that implements a public-facing service for retrieving JSON-formatted shares from the device application; performs database initialisation, storage, and retrieval; and bootstraps the launching of MPC-based operations with other parties and returns the result to the device.
    
    \textbf{\emph{MPyC instance}}. We make use of MPyC---an open-source Python library developed by Schoenmakers~\cite{schoenmakers:mpyc} intended for the rapid prototyping of MPC-based systems. MPyC is based on the VIFF framework by Damg{\aa}rd et al.~\cite{damgaard2009asynchronous}, which implements numerous MPC operation protocols using arithmetic secret sharing, and abstracts the complexity of network management, e.g. socket connections, message formatting and transmission, and reconstructing and computing upon shares between the parties. 
    
    \textbf{\emph{Database}}. Stores time-series IMU shares received from the device application; our implementation uses MongoDB\footnote{MongoDB: \url{https://www.mongodb.com}}, a JSON document-based NoSQL database.


\subsubsection{Deployment}
\label{sec:server_app}

\begin{table}
    \centering
    \caption{MPC server location configurations using AWS.}
    \begin{tabular}{cccc}\toprule
      \textbf{Config} & \textbf{Party 1} & \textbf{Party 2} & \textbf{Party 3} \\\midrule
      \textbf{A} &  \texttt{US\_Oregon} & \texttt{Canada\_Central} & \texttt{US\_Ohio}\\
      \textbf{B}   &  \texttt{US\_Oregon} & \texttt{Canada\_Central} & \texttt{Europe\_London}\\
      \textbf{C}   &  \texttt{US\_Oregon} & \texttt{Europe\_London} & \texttt{Asia\_Singapore} \\\bottomrule
    \end{tabular}
    \label{tab:mpc_aws_configs}
\end{table}

We deploy our architecture using three MPC parties deployed in geographically disparate locations available via AWS: US West (Oregon), US East (Ohio), Central Canada, Europe (London), and South East Asia (Singapore).  These configurations are listed in Table  \ref{tab:mpc_aws_configs}, representing a strictly North American deployment (Config A), North America and Europe (Config B), and North America, Europe and Asia (Config C). The three server-side applications, listed above, were deployed collectively on separate AWS EC2 instances in their respective region; each instance was also pre-loaded with the other instances' TLS certificates and static IP addresses in configuration files.  We use three \texttt{t2.micro} AWS instances, featuring a single-core Intel Xeon Scalable Processor at 3.3 GHz, with 1GB RAM and up to 0.72Gbit/s network throughput. At the time of publication, this incurs an operating cost of \$0.0116 (USD) per hour, or \$8.47 per month per instance~\cite{amazon:ec2}.



\subsubsection{Feature Extraction}
\label{sec:feature_extraction}

In this work, we implement MPC-based feature extraction for replicating the SmartFall proposal by Mauldin et al.~\cite{mauldin2018smartfall}, which uses two features: the samples' vector magnitudes, and the difference between the maximum and minimum magnitudes, $\Delta s$ (Equations \ref{eq:mag} and \ref{eq:delta_s} respectively), over a sliding window of 750ms. This is managed in our framework by aggregating shares in an array of size 24, assuming a sampling frequency of 31.25Hz as in \cite{mauldin2018smartfall}. The vector magnitude is found for each sample via the Newton-Raphson method for computing the square root, while $\Delta s$ relies upon MPyC's maximum and minimum functions of a list of secret-shared elements. These are internally implemented by recursively finding the maximum (or minimum) of each list half, computing their difference, and performing a $\geq$ or $\leq$ operation with zero---the costs of which were listed in Table \ref{tab:comm}.


An important observation is that the choice of features dramatically affects the asymptotic cost of MPC-based operations.  Features reliant upon large numbers of multiplication, comparison or Boolean operations will necessitate many rounds of inter-party communication versus features that can be computed locally, such as additions or multiplications with public values, as noted in Section \ref{sec:mpc_fund}.  We note that the SmartFall features, i.e. the vector magnitude and computing $\Delta s$ in a sliding window, imposes significant cost for the square root and minimum and maximum functions respectively. 

To address this, we also explore derivative-based features, which are used extensively for feature extraction in computer vision tasks \cite{sift, bay_surf, hog}. In particular, inspired by the SURF features used in Bay et al. \cite{bay_surf}, we use the sum of the derivatives and the sum of the squared derivatives for each data channel within the window, resulting in a six-dimensional feature vector, $f$, listed in Equation \ref{eq:f}.

\begin{equation}
    f = \sum \frac{d x}{d t}, \sum \frac{d y}{d t}, \sum \frac{d z}{d t}, \sum \big(\frac{d x}{d t}\big)^{2}, \sum \big(\frac{d y}{d t}\big)^2, \sum \big(\frac{d z}{d t}\big)^2
    \label{eq:f}
\end{equation}

Where $\frac{d x}{d t}$, $\frac{d y}{d t}$, and $\frac{d z}{d t}$ represent the differentials of the $x$, $y$ an $z$ measurement components respectively. The derivative is computed by convolution as follows: 

\begin{equation}
    \frac{d u}{d t} = A_u * h 
\end{equation}

Where $A_u$ is the $u$-th component of an accelerometer measurement and $h$ is a filter kernel with impulse response $[-0.5, 0, 0.5]$.  This feature extraction scheme is substantially more efficient, requiring only a single round of inter-party communication for each derivative squaring operation; the summing operations can be performed locally without communication, as can the convolution operation using a filter kernel that performs multiplications with constant values.

%

\subsubsection{Privacy-Enhancing Inferencing}
\label{sec:inferencing}

We developed a range of MPC-based functions for privacy-preserving inferencing from measurement shares, which were implemented using the fundamental operations from Section~\ref{sec:mpc_fund}.  At present, we have developed MPC-based implementations of the following classifiers.

    \emph{Logistic Regression} (LR) is a linear model of the probability of a binary response given a feature vector, $\vec{x} = [x_1, x_2, \dots, x_n]$.  LR applies the logit function to $\vec{x}$ and parameters, $\vec{\theta}$, that minimise some cost function, e.g. ordinary least squares.  The decision boundary, $h$, is represented by a linear combination of the parameters and features; that is, $h(\vec{\theta}, \vec{x}) = \sum^{n}_{i=0} \theta_{i}x_{i}$.

    
    
    A \emph{Support Vector Machine} (SVM) attempts to find a hyperplane that maximises the margin (distance) between linearly separable data points of classes in the training set.  We implement a traditional SVM with linear kernel for binary classification, whereby samples are classified using a given set of weights, $\vec{w}$, feature vector, $\vec{x}$, bias, $b$, and evaluating Equation~\ref{eq:svm}. Note that, for inferencing,  both SVM and LR reduce to a single inner product. With MPC, a na\"{i}ve approach requires $m$ rounds of communication to compute each $\theta_{i}  x_{i}$; however, \cite{de2012design} demonstrated how to compute the inner product of two vectors using LSSS requiring only a single round of communication with (non-interactive) local summations, which is reproduced in Algorithm \ref{fig:inner_prod} for two shared vectors, $[\textbf{a}]$ and $[\textbf{b}]$.
    
    \begin{equation}
    C=
    \begin{cases}
      1, & \text{if}\ \vec{\theta} \cdot \vec{x} + b \geq 0\\
      0, & \text{otherwise}
    \end{cases}
        \label{eq:svm}
    \end{equation}
    
\begin{algorithm}[t]
\DontPrintSemicolon
\SetAlgoLined
\KwData{$[\textbf{a}] = ([a_1], [a_2],\dots,[a_m]), [\textbf{b}] = ([b_1], [b_2], \dots, [b_m])$}
\KwResult{$[c]$, where $c= \sum^{m}_{j=1}a_j \cdot b_j$}
 \ForEach{$i \in \{1, 2, \dots, 2t+1\}$ \textbf{\textup{in parallel}}}{
    Party $P_i$ computes $d_i \gets \sum^{m}_{j=1}[a_j]_i \cdot [b_j]_i$
    
    $P_i$ shares $d_i$ into $[d_i] \gets SSSS(d_i)$
 }
 $[c] \gets \sum^{2t+1}_{i=1}([d_i] \cdot \prod_{\ell=1, \ell \neq i}^{2t+1}\frac{-\ell}{i-\ell})$
 
 \KwRet $[c]$
 \caption{Inner product with LSSS \cite{chen:ping}.}
 \label{fig:inner_prod}
\end{algorithm}

    \emph{Na\"{i}ve Bayes} (NB) is a probabilistic classifier that uses Bayes' Theorem with a strong independence assumption between features.  A feature vector is subsequently classified by maximising the posterior probability  for  each  class  label using the prior and conditional probabilities computed in the training phase. In our case, we use Gaussian NB to compute the log-likelihoods of real-valued inputs, parameterised by mean $\mu$ with variance $\sigma^{2}$ and evaluated as follows:
    \begin{equation}
    \label{eq:naivebayes}
        g_i(\vec{x}) = \ln(P(C_i)) - \sum_{j=1}^{n}\frac{||x_j-\mu_{ij}||^2}{2\sigma_{ij}^2}
    \end{equation}
    Where $n$ is the number of features and $P(C_i)$ is the probability of the occurrence of class $C_i$.

\section{Evaluation}
\label{sec:eval}

This section presents a two-part evaluation comprising: 1), \emph{classification error rates} achieved by our framework using MPC-based implementations of SmartFall and derivative-based features with the three inferencing algorithms; and 2), \emph{computational performance}, for evaluating the time complexity of the framework. 






\subsection{Classification Error Rates}


We now present error rates achieved by the models used in our MPC-based implementations of SmartFall~\cite{mauldin2018smartfall} using the standard and derivative-based features described in Section \ref{sec:feature_extraction}.  The authors of \cite{mauldin2018smartfall} provide a publicly-available dataset comprising labelled IMU accelerometer measurements. These correspond to binary fall events from seven participants between 21--55 years old wearing a Microsoft Band 2 smartwatch. 

An off-line Python application was used to generate the model parameters.  Firstly, a sliding window of 750ms was used with data sampled at a frequency of 31.25Hz, over which the vector magnitude and $\Delta s$ were computed. This was repeated for the derivative-based feature extraction.  Next, six classifiers---Gaussian NB, SVM (with linear kernel), and logistic regression models for both feature types---were trained independently using the Scikit-Learn library~\cite{pedregosa2011scikit}. The inclusion of logistic regression goes beyond the classifiers evaluated in \cite{mauldin2018smartfall}. The dataset was split using a 80:20 training-test set ratio, and five-fold cross-validation was used to select the best performing model, whose final error rate was measured using the test set.  The parameters of the best performing models were subsequently exported and implemented within the inferencing procedures detailed in Section \ref{sec:inferencing}. 

Tables \ref{tab:errors} and \ref{tab:errors_gradient} presents the error rates for each model with respect to classification accuracy, precision, recall, and F1-scores averaged across all participants for the SmartFall and derivative-based features respectively.  The achieved results, i.e. $93.6\%$ accuracy and $0.733$ F1-score in the best case for SVM, are commensurate with the original SmartFall proposal and other fall detection work listed previously in Table \ref{tab:related}. We note that the derivative features outperform the original features in terms of precision, recall and F1-score, e.g. $0.804$ F1-score for NB versus $0.732$, and $0.835$ recall for SVM versus $0.663$; however, the classification accuracy is broadly similar for both feature types.

\begin{table}
\centering
\caption{Classification error with SmartFall features.}
\begin{threeparttable}
\begin{tabular}{@{}ccccc@{}}
\toprule
\textbf{Algorithm} & \textbf{Accuracy} & \textbf{Precision} & \textbf{Recall} & \textbf{F1-Score}\\ \midrule
\textbf{LR} & 0.933 & 0.780 & 0.688 & 0.731\\
\textbf{NB} & 0.926 & 0.704 & 0.762 & 0.732\\
\textbf{SVM} & 0.936 & 0.819 & 0.663 & 0.733\\\bottomrule
\end{tabular}
\begin{tablenotes}
\item LR: Logistic Regression, NB: Na\"{i}ve Bayes, SVM: Support Vector Machine.
\end{tablenotes}
\end{threeparttable}
\label{tab:errors}
\end{table}
\begin{table}
\centering
\caption{Classification error with derivative-based features.}
\begin{threeparttable}
\begin{tabular}{@{}ccccc@{}}
\toprule
\textbf{Algorithm} & \textbf{Accuracy} & \textbf{Precision} & \textbf{Recall} & \textbf{F1-Score}\\ \midrule
\textbf{LR} & 0.936 & 0.771 & 0.812 & 0.791\\
\textbf{NB} & 0.934 & 0.725 & 0.903 & 0.804\\
\textbf{SVM} & 0.937 & 0.762 & 0.835 & 0.797\\\bottomrule
\end{tabular}
\end{threeparttable}
\label{tab:errors_gradient}
\end{table}

\subsection{Computational Performance}
 This suite of experiments evaluates the latencies incurred by the device and cloud entities. 
\subsubsection{Device-side Latency}
Here, we measure the average time to both construct the shares using SSSS and their transmission time to $n=3$ parties, with a reconstruction threshold of $t=3$.  This is measured for the sharing and transmission of samples in a sliding window of 750ms (21 samples at 31.25Hz).  The mean share construction time, averaged across all windows in the dataset, took 6.1ms ($\sigma=0.9$), with the upload time taking 398ms ($\sigma=23.9$ms), 404ms ($\sigma=25.9$ms) and 493.2ms ($\sigma=33.9$ms) for configurations A, B and C respectively. These timings are measured by the device-side Python application using the \texttt{time} module, which executed upon a Raspberry Pi 3 located in Belgium, Europe. The device specifications were listed previously in Section \ref{sec:device_app}.



\begin{table}
\centering
\caption{Mean feature extraction and inferencing times (milliseconds) using SmartFall features for each configuration; S.D. shown in brackets.}
\resizebox{\linewidth}{!}{%
\begin{threeparttable}
\begin{tabular}{@{}ccccccc@{}}
\toprule
\multicolumn{1}{l}{} & \multicolumn{2}{c}{\textbf{Config A}} & \multicolumn{2}{c}{\textbf{Config B}} & \multicolumn{2}{c}{\textbf{Config C}} \\ \midrule
\textbf{Alg.} & \textbf{FE} & \textbf{IN} & \textbf{FE} & \textbf{IN} & \textbf{FE} & \textbf{IN} \\\midrule
\textbf{LR} & | & 308 (91) & | & 383 (104) & | &  507 (150) \\
\textbf{SVM} & \textbf{779 (119)}$^{\dag}$  & --- &  \textbf{1005 (125)} & --- &  \textbf{1204 (194)} & --- \\
\textbf{NB} & | & 509 (103) & | & 557 (226) & | &  692 (114) \\ \bottomrule
\end{tabular}
\begin{tablenotes}
\item $\dag$: Common feature extraction for all algorithms,  | : Repeat all column-wise entries, ---: Repeat previous entry (column-wise), Alg.: Algorithm, FE: Feature extraction, IN: Inferencing.
\end{tablenotes}
\end{threeparttable}
}
\label{tab:smartfall_times}
\end{table}

\begin{table}
\centering
\caption{Mean feature extraction and inferencing times (milliseconds) using derivative-based features.}
\resizebox{0.92\linewidth}{!}{%
\begin{threeparttable}
\begin{tabular}{@{}ccccccc@{}}
\toprule
\multicolumn{1}{l}{} & \multicolumn{2}{c}{\textbf{Config A}} & \multicolumn{2}{c}{\textbf{Config B}} & \multicolumn{2}{c}{\textbf{Config C}} \\ \midrule
\textbf{Alg.} & \textbf{FE} & \textbf{IN} & \textbf{FE} & \textbf{IN} & \textbf{FE} & \textbf{IN} \\\midrule
\textbf{LR} & | & 297 (106) & | & 365 (95) &   | &  490 (104) \\
\textbf{SVM} & \textbf{68 (15)}$^{\dag}$ & --- &  \textbf{102 (19)} & --- & \textbf{111 (24)} & --- \\
\textbf{NB} & | & 552 (103) & | &  602 (106) & | &  742 (113) \\ \bottomrule
\end{tabular}
\begin{tablenotes}
\item  $\dag$: Common feature extraction for all algorithms, | : Repeat all column-wise entries, ---: Repeat previous entry (column-wise), Alg.: Algorithm, FE: Feature extraction, IN: Inferencing.
\end{tablenotes}
\end{threeparttable}
}
\label{tab:gradient_times}
\end{table}

\subsubsection{Cloud-side Latency}

\begin{figure}
    \centering
    \includegraphics[width=0.86\linewidth]{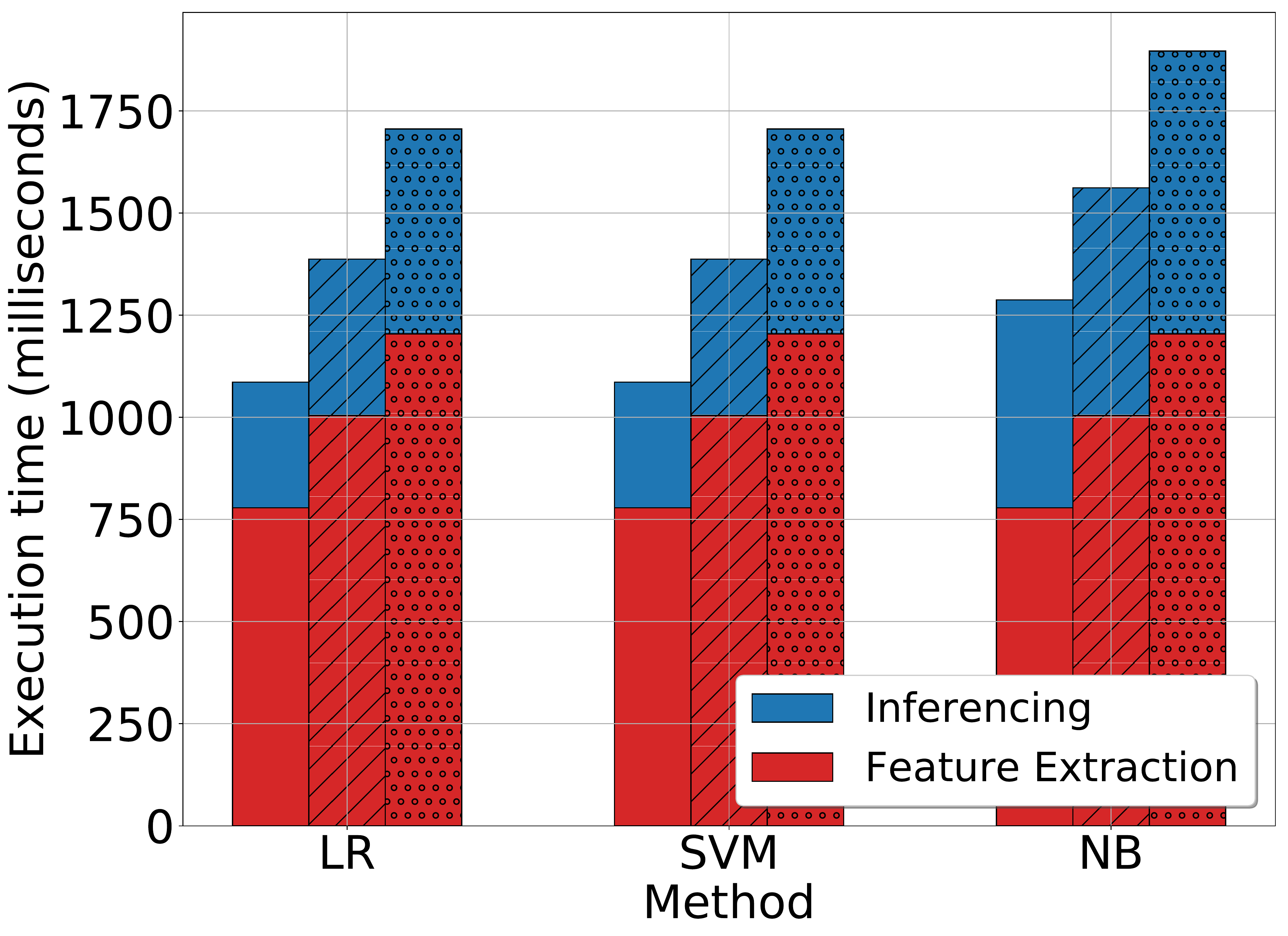}
    \caption{Mean window execution times using SmartFall features for Config. A (no hatch), B (hatched), and C (dotted).}
    \label{fig:execution_times_smartfall}
\end{figure}

\begin{figure}
    \centering
    \includegraphics[width=0.88\linewidth]{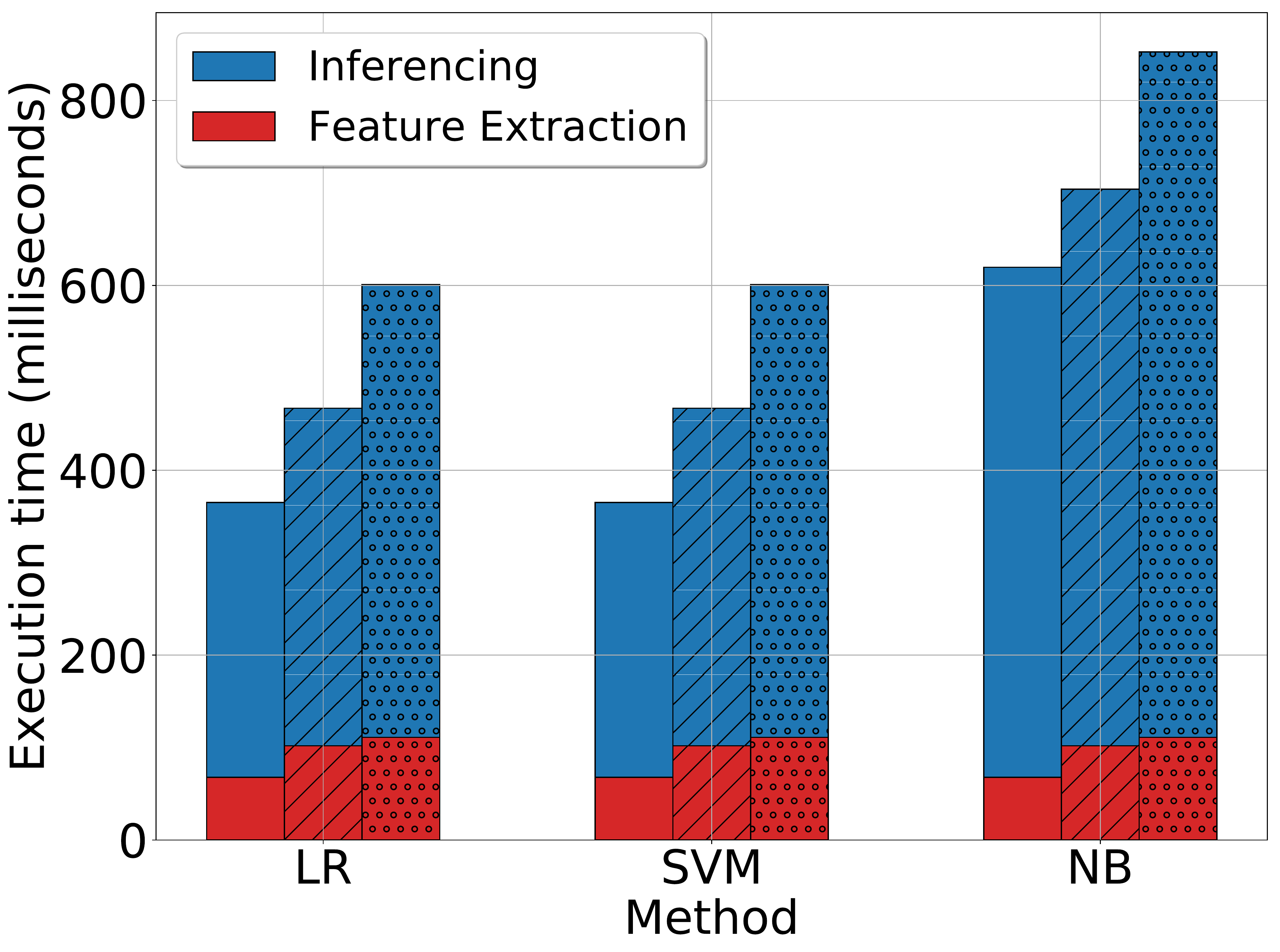}
    \caption{Mean window execution times using derivative features for Config. A (no hatch), B (hatched), and C (dotted).}
    \label{fig:execution_times_derivative}
\end{figure}

We measure the average wall-clock times of the MPC-based feature extraction and inferencing procedures, which were described in Sections \ref{sec:feature_extraction} and \ref{sec:inferencing}. The total time for classifying each window can be considered as the sum of these two values. These were measured within the server-side Python application also via the \texttt{time} module.  We report the times for each cloud configuration using the standard SmartFall features in Table \ref{tab:smartfall_times} and the derivative-based features in Table \ref{tab:gradient_times}. The results are also illustrated in Figures \ref{fig:execution_times_smartfall} and \ref{fig:execution_times_derivative}.    


\subsection{Discussion}

One concern with MPC, raised previously in Section~\ref{sec:mpc_fund}, is the inter-party communication cost for performing multiplication- and comparison-heavy functions, which were used in the feature extraction and inferencing procedures.  With the original SmartFall features, inferencing and feature extraction takes approximately 1900ms in the worst case with Naive Bayes under Config C, to $\sim$1100ms in the best case using LR with Config A. We also observe that the feature extraction stage dominates the execution times using this feature type, accounting for approximately two-thirds of each tested algorithm's total time.  In contrast, this inverts with the use of derivative features, which require fewer rounds of inter-party communication; the \emph{inferencing} stage becomes the dominant factor for all algorithms, rather than the feature extraction.  The execution time is also substantially reduced for the derivative-based features; reducing from 1087ms to 365ms using LR with Config A, for example, between the SmartFall and derivative-based features respectively.  In general, the derivative-based features are extracted $\sim$11x faster than the originally proposed SmartFall features in \cite{mauldin2018smartfall}; the \emph{total} execution time, including inferencing, exhibits a $\sim$2.8x increase between derivative- and SmartFall-based experiments.
 
We also see that the SVM and LR algorithms execute in similar times due to the mathematical similarities of inferencing, requiring an inner product operation that necessitates only a single round of communication.  NB, however, is relatively expensive from requiring two multiplications for each feature in the window: one for finding the square of the feature subtracted from the mean, and the inverse multiplication (division) of this value by the variance.  The choice of server configuration, i.e. the parties' geographical distribution, also has a significant effect on the total execution time; increasing, for example, from approximately 1087ms to 1711ms using SmartFall features with SVM-based inferencing, corresponding to a $\sim$57\% increase from Config A to C. Similar results are exhibited for all SmartFall- and derivative-based experiments.  

Lastly, we observe that, in some instances, the total execution time falls within that required for acquiring the measurements within the window (750ms). This applies for all derivative-based prediction using configurations A and B, but only the SVM and LR inferencing algorithms using Config C. This leads us to believe that these would be best suited for real-time prediction.

\section{Conclusion}
\label{sec:conc}
A plethora of fall detection systems rely upon IMU data from user-owned commodity devices.  However, as we discussed in Section \ref{sec:fall_detection}, these proposals necessitate the collection of data that can be used to identify users' gait, their position using dead reckoning, and determining their activities of daily living (ADLs).  In light of this, we presented the first investigation of the application of multi-party computation for IMU-based fall detection with the aim of preserving the confidentiality of sensor measurements.

We presented the design and implementation of our system in Sections \ref{sec:private_falldetection} and \ref{sec:imp}, which was subsequently deployed using three AWS EC2 instances in geographically disparate locations under three configurations. Each instance acted as an MPC participant for jointly computing the feature extraction and inferencing procedures using weights from pre-trained global models.  The system was evaluated with the original SmartFall features alongside an exploration of derivative-based features for reducing the communication complexity imposed by MPC operations.

In Section \ref{sec:eval}, we presented the results from a two-part evaluation of the system's classification error and computational performance. This was conducted using a publicly-available dataset comprising data from seven participants wearing an off-the-shelf device. The performance results indicate that state-of-the-art error rates can be achieved for MPC-based fall detection---achieving 0.926--0.937 classification accuracy and 0.731--0.804 F1-score depending on the chosen algorithm and feature set. 

We also showed experimentally that the choices of inferencing algorithm, feature extraction, and server location have substantial effects on computational performance. The worst cases increase the total execution total by almost three-fold versus the best performers, which is dominated principally by the choice of features. In the best cases---for example, using an SVM classifier with derivative-based features in a cloud configuration in the same continent---the total prediction time is less than the on-device data acquisition time. This provides an indication that real-time prediction is feasible. 

To conclude, we aim to pursue the following research directions in future work:
\begin{itemize}
    \item \emph{MPC-based fall detection using deep learning}. Existing fall detection systems rely heavily on traditional machine learning classifiers; a small selection of recent work, however, has begun to explore the use of deep neural networks (DNNs), such as RNNs~\cite{musci2018online}. As such, a future avenue lies in exploring MPC-based DNNs and the overhead imposed by evaluating large numbers of intermediate layers with potentially hundreds or thousands of nodes.
    \item \emph{Video-based fall detection}. A separate paradigm of fall detection systems, described in Section \ref{sec:fall_detection}, relies upon video camera data from devices situated in the user's ambient environment. This data, however, is of significantly greater dimension than the IMU data explored in this work; a future direction is investigating the extent to which MPC-based fall detection can be performed with high-frequency, high-dimensionality data.

\end{itemize}

\balance


\bibliographystyle{ACM-Reference-Format}
\bibliography{refs}


\begin{thebibliography}{49}


\ifx \showCODEN    \undefined \def \showCODEN     #1{\unskip}     \fi
\ifx \showDOI      \undefined \def \showDOI       #1{#1}\fi
\ifx \showISBNx    \undefined \def \showISBNx     #1{\unskip}     \fi
\ifx \showISBNxiii \undefined \def \showISBNxiii  #1{\unskip}     \fi
\ifx \showISSN     \undefined \def \showISSN      #1{\unskip}     \fi
\ifx \showLCCN     \undefined \def \showLCCN      #1{\unskip}     \fi
\ifx \shownote     \undefined \def \shownote      #1{#1}          \fi
\ifx \showarticletitle \undefined \def \showarticletitle #1{#1}   \fi
\ifx \showURL      \undefined \def \showURL       {\relax}        \fi
\providecommand\bibfield[2]{#2}
\providecommand\bibinfo[2]{#2}
\providecommand\natexlab[1]{#1}
\providecommand\showeprint[2][]{arXiv:#2}

\bibitem[\protect\citeauthoryear{Albert, Kording, Herrmann, and
  Jayaraman}{Albert et~al\mbox{.}}{2012}]%
        {albert2012fall}
\bibfield{author}{\bibinfo{person}{Mark~V Albert}, \bibinfo{person}{Konrad
  Kording}, \bibinfo{person}{Megan Herrmann}, {and} \bibinfo{person}{Arun
  Jayaraman}.} \bibinfo{year}{2012}\natexlab{}.
\newblock \showarticletitle{Fall classification by machine learning using
  mobile phones}.
\newblock \bibinfo{journal}{\emph{PloS one}} \bibinfo{volume}{7},
  \bibinfo{number}{5} (\bibinfo{year}{2012}).
\newblock


\bibitem[\protect\citeauthoryear{Amazon}{Amazon}{2019}]%
        {amazon:ec2}
\bibfield{author}{\bibinfo{person}{Amazon}.} \bibinfo{year}{2019}\natexlab{}.
\newblock \bibinfo{title}{{Amazon EC2 T2} instances}.
\newblock
\newblock
\newblock
\shownote{\url{https://aws.amazon.com/ec2/instance-types/t2/}.}


\bibitem[\protect\citeauthoryear{Bay, Ess, Tuytelaars, and Van~Gool}{Bay
  et~al\mbox{.}}{2008}]%
        {bay_surf}
\bibfield{author}{\bibinfo{person}{Herbert Bay}, \bibinfo{person}{Andreas Ess},
  \bibinfo{person}{Tinne Tuytelaars}, {and} \bibinfo{person}{Luc Van~Gool}.}
  \bibinfo{year}{2008}\natexlab{}.
\newblock \showarticletitle{Speeded-Up Robust Features (SURF)}.
\newblock \bibinfo{journal}{\emph{Comput. Vis. Image Underst.}}
  \bibinfo{volume}{110}, \bibinfo{number}{3} (\bibinfo{date}{June}
  \bibinfo{year}{2008}), \bibinfo{pages}{346--359}.
\newblock


\bibitem[\protect\citeauthoryear{Ben-Or, Goldwasser, and Wigderson}{Ben-Or
  et~al\mbox{.}}{1988}]%
        {ben1988completeness}
\bibfield{author}{\bibinfo{person}{Michael Ben-Or}, \bibinfo{person}{Shafi
  Goldwasser}, {and} \bibinfo{person}{Avi Wigderson}.}
  \bibinfo{year}{1988}\natexlab{}.
\newblock \showarticletitle{Completeness theorems for non-cryptographic
  fault-tolerant distributed computation}. In
  \bibinfo{booktitle}{\emph{Proceedings of the 20th Annual ACM Symposium on
  Theory of Computing}}. ACM.
\newblock


\bibitem[\protect\citeauthoryear{Cao, Yang, and Liu}{Cao et~al\mbox{.}}{2012}]%
        {cao2012falld}
\bibfield{author}{\bibinfo{person}{Yabo Cao}, \bibinfo{person}{Yujiu Yang},
  {and} \bibinfo{person}{WenHuang Liu}.} \bibinfo{year}{2012}\natexlab{}.
\newblock \showarticletitle{E-FallD: A fall detection system using
  android-based smartphone}. In \bibinfo{booktitle}{\emph{9th International
  Conference on Fuzzy Systems and Knowledge Discovery}}. IEEE,
  \bibinfo{pages}{1509--1513}.
\newblock


\bibitem[\protect\citeauthoryear{Chen}{Chen}{2012}]%
        {chen:ping}
\bibfield{author}{\bibinfo{person}{Ping Chen}.}
  \bibinfo{year}{2012}\natexlab{}.
\newblock \emph{\bibinfo{title}{Secure Multiparty Computation for Privacy
  Preserving Data Mining}}.
\newblock \bibinfo{thesistype}{Master's\ thesis}. \bibinfo{school}{TU
  Eindhoven}.
\newblock


\bibitem[\protect\citeauthoryear{Continella, Polino, Pogliani, and
  Zanero}{Continella et~al\mbox{.}}{2018}]%
        {continella2018there}
\bibfield{author}{\bibinfo{person}{Andrea Continella}, \bibinfo{person}{Mario
  Polino}, \bibinfo{person}{Marcello Pogliani}, {and} \bibinfo{person}{Stefano
  Zanero}.} \bibinfo{year}{2018}\natexlab{}.
\newblock \showarticletitle{There's a hole in that bucket!: A large-scale
  analysis of misconfigured {S3} buckets}. In
  \bibinfo{booktitle}{\emph{Proceedings of the 34th Annual Computer Security
  Applications Conference}}. ACM, \bibinfo{pages}{702--711}.
\newblock


\bibitem[\protect\citeauthoryear{Dalal and Triggs}{Dalal and Triggs}{2005}]%
        {hog}
\bibfield{author}{\bibinfo{person}{Navneet Dalal} {and} \bibinfo{person}{Bill
  Triggs}.} \bibinfo{year}{2005}\natexlab{}.
\newblock \showarticletitle{Histograms of oriented gradients for human
  detection}. In \bibinfo{booktitle}{\emph{Proceedings of IEEE Computer Society
  Conference on Computer Vision and Pattern Recognition}}.
  \bibinfo{publisher}{IEEE Computer Society}, \bibinfo{pages}{886--893}.
\newblock


\bibitem[\protect\citeauthoryear{Damg{\aa}rd, Geisler, Kr{\o}igaard, and
  Nielsen}{Damg{\aa}rd et~al\mbox{.}}{2009}]%
        {damgaard2009asynchronous}
\bibfield{author}{\bibinfo{person}{Ivan Damg{\aa}rd}, \bibinfo{person}{Martin
  Geisler}, \bibinfo{person}{Mikkel Kr{\o}igaard}, {and}
  \bibinfo{person}{Jesper~Buus Nielsen}.} \bibinfo{year}{2009}\natexlab{}.
\newblock \showarticletitle{Asynchronous multiparty computation: Theory and
  implementation}. In \bibinfo{booktitle}{\emph{International Workshop on
  Public Key Cryptography}}. Springer, \bibinfo{pages}{160--179}.
\newblock


\bibitem[\protect\citeauthoryear{de~Hoogh}{de~Hoogh}{2012}]%
        {de2012design}
\bibfield{author}{\bibinfo{person}{Sebastiaan de Hoogh}.}
  \bibinfo{year}{2012}\natexlab{}.
\newblock \showarticletitle{Design of large scale applications of secure
  multiparty computation: secure linear programming}.
\newblock \bibinfo{journal}{\emph{Ph. D. dissertation}} (\bibinfo{year}{2012}).
\newblock


\bibitem[\protect\citeauthoryear{de~Hoogh, Schoenmakers, Chen, and op~den
  Akker}{de~Hoogh et~al\mbox{.}}{2014}]%
        {de2014practical}
\bibfield{author}{\bibinfo{person}{Sebastiaan de Hoogh}, \bibinfo{person}{Berry
  Schoenmakers}, \bibinfo{person}{Ping Chen}, {and} \bibinfo{person}{Harm
  op~den Akker}.} \bibinfo{year}{2014}\natexlab{}.
\newblock \showarticletitle{Practical secure decision tree learning in a
  teletreatment application}. In \bibinfo{booktitle}{\emph{International
  Conference on Financial Cryptography and Data Security}}. Springer.
\newblock


\bibitem[\protect\citeauthoryear{Delahoz and Labrador}{Delahoz and
  Labrador}{2014}]%
        {delahoz2014survey}
\bibfield{author}{\bibinfo{person}{Yueng~Santiago Delahoz} {and}
  \bibinfo{person}{Miguel~Angel Labrador}.} \bibinfo{year}{2014}\natexlab{}.
\newblock \showarticletitle{Survey on fall detection and fall prevention using
  wearable and external sensors}.
\newblock \bibinfo{journal}{\emph{Sensors}} \bibinfo{volume}{14},
  \bibinfo{number}{10} (\bibinfo{year}{2014}), \bibinfo{pages}{19806--19842}.
\newblock


\bibitem[\protect\citeauthoryear{Doukas and Maglogiannis}{Doukas and
  Maglogiannis}{2011}]%
        {doukas2011managing}
\bibfield{author}{\bibinfo{person}{Charalampos Doukas} {and}
  \bibinfo{person}{Ilias Maglogiannis}.} \bibinfo{year}{2011}\natexlab{}.
\newblock \showarticletitle{Managing wearable sensor data through cloud
  computing}. In \bibinfo{booktitle}{\emph{3rd IEEE International Conference on
  Cloud Computing Technology and Science}}. IEEE.
\newblock


\bibitem[\protect\citeauthoryear{Doukas, Maglogiannis, Tragas, Liapis, and
  Yovanof}{Doukas et~al\mbox{.}}{2007}]%
        {doukas2007patient}
\bibfield{author}{\bibinfo{person}{Charalampos Doukas}, \bibinfo{person}{Ilias
  Maglogiannis}, \bibinfo{person}{Philippos Tragas}, \bibinfo{person}{Dimitris
  Liapis}, {and} \bibinfo{person}{Gregory Yovanof}.}
  \bibinfo{year}{2007}\natexlab{}.
\newblock \showarticletitle{Patient fall detection using support vector
  machines}. In \bibinfo{booktitle}{\emph{IFIP International Conference on
  Artificial Intelligence Applications and Innovations}}. Springer,
  \bibinfo{pages}{147--156}.
\newblock


\bibitem[\protect\citeauthoryear{Edgcomb and Vahid}{Edgcomb and Vahid}{2012}]%
        {edgcomb2012automated}
\bibfield{author}{\bibinfo{person}{Alex Edgcomb} {and} \bibinfo{person}{Frank
  Vahid}.} \bibinfo{year}{2012}\natexlab{}.
\newblock \showarticletitle{Automated fall detection on privacy-enhanced
  video}. In \bibinfo{booktitle}{\emph{Engineering in Medicine and Biology
  Society (EMBC), Annual International Conference of the IEEE}}. IEEE,
  \bibinfo{pages}{252--255}.
\newblock


\bibitem[\protect\citeauthoryear{Florence, Bergen, Atherly, Burns, Stevens, and
  Drake}{Florence et~al\mbox{.}}{2018}]%
        {florence2018medical}
\bibfield{author}{\bibinfo{person}{Curtis~S Florence}, \bibinfo{person}{Gwen
  Bergen}, \bibinfo{person}{Adam Atherly}, \bibinfo{person}{Elizabeth Burns},
  \bibinfo{person}{Judy Stevens}, {and} \bibinfo{person}{Cynthia Drake}.}
  \bibinfo{year}{2018}\natexlab{}.
\newblock \showarticletitle{Medical costs of fatal and nonfatal falls in older
  adults}.
\newblock \bibinfo{journal}{\emph{Journal of the American Geriatrics Society}}
  \bibinfo{volume}{66}, \bibinfo{number}{4} (\bibinfo{year}{2018}),
  \bibinfo{pages}{693--698}.
\newblock


\bibitem[\protect\citeauthoryear{Fortino, Parisi, Pirrone, and
  Di~Fatta}{Fortino et~al\mbox{.}}{2014}]%
        {fortino2014bodycloud}
\bibfield{author}{\bibinfo{person}{Giancarlo Fortino}, \bibinfo{person}{Daniele
  Parisi}, \bibinfo{person}{Vincenzo Pirrone}, {and} \bibinfo{person}{Giuseppe
  Di~Fatta}.} \bibinfo{year}{2014}\natexlab{}.
\newblock \showarticletitle{BodyCloud: A SaaS approach for community body
  sensor networks}.
\newblock \bibinfo{journal}{\emph{Future Generation Computer Systems}}
  \bibinfo{volume}{35} (\bibinfo{year}{2014}), \bibinfo{pages}{62--79}.
\newblock


\bibitem[\protect\citeauthoryear{Gafurov}{Gafurov}{2007}]%
        {gafurov2007survey}
\bibfield{author}{\bibinfo{person}{Davrondzhon Gafurov}.}
  \bibinfo{year}{2007}\natexlab{}.
\newblock \showarticletitle{A survey of biometric gait recognition: Approaches,
  security and challenges}. In \bibinfo{booktitle}{\emph{Annual Norwegian
  Computer Science Conference}}.
\newblock


\bibitem[\protect\citeauthoryear{Gafurov, Snekkenes, and Bours}{Gafurov
  et~al\mbox{.}}{2007}]%
        {gafurov2007gait}
\bibfield{author}{\bibinfo{person}{Davrondzhon Gafurov}, \bibinfo{person}{Einar
  Snekkenes}, {and} \bibinfo{person}{Patrick Bours}.}
  \bibinfo{year}{2007}\natexlab{}.
\newblock \showarticletitle{Gait authentication and identification using
  wearable accelerometer sensor}. In \bibinfo{booktitle}{\emph{IEEE Workshop on
  Automatic Identification Advanced Technologies}}. IEEE,
  \bibinfo{pages}{220--225}.
\newblock


\bibitem[\protect\citeauthoryear{Gravina, Ma, Pace, Aloi, Russo, Li, and
  Fortino}{Gravina et~al\mbox{.}}{2017}]%
        {gravina2017cloud}
\bibfield{author}{\bibinfo{person}{Raffaele Gravina}, \bibinfo{person}{Congcong
  Ma}, \bibinfo{person}{Pasquale Pace}, \bibinfo{person}{Gianluca Aloi},
  \bibinfo{person}{Wilma Russo}, \bibinfo{person}{Wenfeng Li}, {and}
  \bibinfo{person}{Giancarlo Fortino}.} \bibinfo{year}{2017}\natexlab{}.
\newblock \showarticletitle{Cloud-based Activity-aaService cyber-physical
  framework for human activity monitoring in mobility}.
\newblock \bibinfo{journal}{\emph{Future Generation Computer Systems}}
  \bibinfo{volume}{75} (\bibinfo{year}{2017}), \bibinfo{pages}{158--171}.
\newblock


\bibitem[\protect\citeauthoryear{Hong, Kim, Ahn, and Kim}{Hong
  et~al\mbox{.}}{2010}]%
        {hong2010mobile}
\bibfield{author}{\bibinfo{person}{Yu-Jin Hong}, \bibinfo{person}{Ig-Jae Kim},
  \bibinfo{person}{Sang~Chul Ahn}, {and} \bibinfo{person}{Hyoung-Gon Kim}.}
  \bibinfo{year}{2010}\natexlab{}.
\newblock \showarticletitle{Mobile health monitoring system based on activity
  recognition using accelerometer}.
\newblock \bibinfo{journal}{\emph{Simulation Modelling Practice and Theory}}
  \bibinfo{volume}{18}, \bibinfo{number}{4} (\bibinfo{year}{2010}),
  \bibinfo{pages}{446--455}.
\newblock


\bibitem[\protect\citeauthoryear{Juefei-Xu, Bhagavatula, Jaech, Prasad, and
  Savvides}{Juefei-Xu et~al\mbox{.}}{2012}]%
        {juefei2012gait}
\bibfield{author}{\bibinfo{person}{Felix Juefei-Xu},
  \bibinfo{person}{Chandrasekhar Bhagavatula}, \bibinfo{person}{Aaron Jaech},
  \bibinfo{person}{Unni Prasad}, {and} \bibinfo{person}{Marios Savvides}.}
  \bibinfo{year}{2012}\natexlab{}.
\newblock \showarticletitle{Gait-ID on the move: Pace independent human
  identification using cell phone accelerometer dynamics}. In
  \bibinfo{booktitle}{\emph{Fifth International Conference on Biometrics:
  Theory, Applications and Systems}} \emph{(\bibinfo{series}{BTAS})}. IEEE,
  \bibinfo{pages}{8--15}.
\newblock


\bibitem[\protect\citeauthoryear{Kang and Han}{Kang and Han}{2015}]%
        {kang2015smartpdr}
\bibfield{author}{\bibinfo{person}{Wonho Kang} {and} \bibinfo{person}{Youngnam
  Han}.} \bibinfo{year}{2015}\natexlab{}.
\newblock \showarticletitle{SmartPDR: Smartphone-based pedestrian dead
  reckoning for indoor localization}.
\newblock \bibinfo{journal}{\emph{IEEE Sensors Journal}} \bibinfo{volume}{15},
  \bibinfo{number}{5} (\bibinfo{year}{2015}).
\newblock


\bibitem[\protect\citeauthoryear{Kwolek and Kepski}{Kwolek and Kepski}{2014}]%
        {kwolek2014human}
\bibfield{author}{\bibinfo{person}{Bogdan Kwolek} {and} \bibinfo{person}{Michal
  Kepski}.} \bibinfo{year}{2014}\natexlab{}.
\newblock \showarticletitle{Human fall detection on embedded platform using
  depth maps and wireless accelerometer}.
\newblock \bibinfo{journal}{\emph{Computer methods and programs in
  biomedicine}} \bibinfo{volume}{117}, \bibinfo{number}{3}
  (\bibinfo{year}{2014}), \bibinfo{pages}{489--501}.
\newblock


\bibitem[\protect\citeauthoryear{Liu and Cheng}{Liu and Cheng}{2012}]%
        {liu2012fall}
\bibfield{author}{\bibinfo{person}{Shing-Hong Liu} {and}
  \bibinfo{person}{Wen-Chang Cheng}.} \bibinfo{year}{2012}\natexlab{}.
\newblock \showarticletitle{Fall detection with the support vector machine
  during scripted and continuous unscripted activities}.
\newblock \bibinfo{journal}{\emph{Sensors}} \bibinfo{volume}{12},
  \bibinfo{number}{9} (\bibinfo{year}{2012}), \bibinfo{pages}{12301--12316}.
\newblock


\bibitem[\protect\citeauthoryear{Lowe}{Lowe}{2004}]%
        {sift}
\bibfield{author}{\bibinfo{person}{David~G. Lowe}.}
  \bibinfo{year}{2004}\natexlab{}.
\newblock \showarticletitle{Distinctive image features from scale-invariant
  keypoints}.
\newblock \bibinfo{journal}{\emph{Int. J. Comput. Vision}}
  \bibinfo{volume}{60}, \bibinfo{number}{2} (\bibinfo{date}{Nov.}
  \bibinfo{year}{2004}), \bibinfo{pages}{91--110}.
\newblock


\bibitem[\protect\citeauthoryear{Lu{\v{s}}trek and Kalu{\v{z}}a}{Lu{\v{s}}trek
  and Kalu{\v{z}}a}{2009}]%
        {luvstrek2009fall}
\bibfield{author}{\bibinfo{person}{Mitja Lu{\v{s}}trek} {and}
  \bibinfo{person}{Bo{\v{s}}tjan Kalu{\v{z}}a}.}
  \bibinfo{year}{2009}\natexlab{}.
\newblock \showarticletitle{Fall detection and activity recognition with
  machine learning}.
\newblock \bibinfo{journal}{\emph{Informatica}} \bibinfo{volume}{33},
  \bibinfo{number}{2} (\bibinfo{year}{2009}).
\newblock


\bibitem[\protect\citeauthoryear{Makri, Rotaru, Smart, and Vercauteren}{Makri
  et~al\mbox{.}}{2017}]%
        {epic}
\bibfield{author}{\bibinfo{person}{Eleftheria Makri}, \bibinfo{person}{Dragos
  Rotaru}, \bibinfo{person}{Nigel~P. Smart}, {and} \bibinfo{person}{Frederik
  Vercauteren}.} \bibinfo{year}{2017}\natexlab{}.
\newblock \bibinfo{title}{EPIC: Efficient Private Image Classification (or:
  Learning from the Masters)}.
\newblock \bibinfo{howpublished}{Cryptology ePrint Archive, Report 2017/1190}.
\newblock
\newblock
\shownote{\url{https://eprint.iacr.org/2017/1190}.}


\bibitem[\protect\citeauthoryear{Mantyjarvi, Lindholm, Vildjiounaite, Makela,
  and Ailisto}{Mantyjarvi et~al\mbox{.}}{2005}]%
        {mantyjarvi2005identifying}
\bibfield{author}{\bibinfo{person}{Jani Mantyjarvi}, \bibinfo{person}{Mikko
  Lindholm}, \bibinfo{person}{Elena Vildjiounaite}, \bibinfo{person}{S-M
  Makela}, {and} \bibinfo{person}{HA Ailisto}.}
  \bibinfo{year}{2005}\natexlab{}.
\newblock \showarticletitle{Identifying users of portable devices from gait
  pattern with accelerometers}. In \bibinfo{booktitle}{\emph{IEEE International
  Conference on Acoustics, Speech, and Signal Processing}}. IEEE.
\newblock


\bibitem[\protect\citeauthoryear{Mastorakis and Makris}{Mastorakis and
  Makris}{2014}]%
        {mastorakis2014fall}
\bibfield{author}{\bibinfo{person}{Georgios Mastorakis} {and}
  \bibinfo{person}{Dimitrios Makris}.} \bibinfo{year}{2014}\natexlab{}.
\newblock \showarticletitle{Fall detection system using Kinect's infrared
  sensor}.
\newblock \bibinfo{journal}{\emph{Journal of Real-Time Image Processing}}
  \bibinfo{volume}{9}, \bibinfo{number}{4} (\bibinfo{year}{2014}).
\newblock


\bibitem[\protect\citeauthoryear{Mauldin, Canby, Metsis, Ngu, and
  Rivera}{Mauldin et~al\mbox{.}}{2018}]%
        {mauldin2018smartfall}
\bibfield{author}{\bibinfo{person}{Taylor Mauldin}, \bibinfo{person}{Marc
  Canby}, \bibinfo{person}{Vangelis Metsis}, \bibinfo{person}{Anne Ngu}, {and}
  \bibinfo{person}{Coralys Rivera}.} \bibinfo{year}{2018}\natexlab{}.
\newblock \showarticletitle{SmartFall: a smartwatch-based fall detection system
  using deep learning}.
\newblock \bibinfo{journal}{\emph{Sensors}} \bibinfo{volume}{18},
  \bibinfo{number}{10} (\bibinfo{year}{2018}), \bibinfo{pages}{3363}.
\newblock


\bibitem[\protect\citeauthoryear{Mellone, Tacconi, Schwickert, Klenk, Becker,
  and Chiari}{Mellone et~al\mbox{.}}{2012}]%
        {mellone2012smartphone}
\bibfield{author}{\bibinfo{person}{S Mellone}, \bibinfo{person}{C Tacconi},
  \bibinfo{person}{L Schwickert}, \bibinfo{person}{J Klenk}, \bibinfo{person}{C
  Becker}, {and} \bibinfo{person}{L Chiari}.} \bibinfo{year}{2012}\natexlab{}.
\newblock \showarticletitle{Smartphone-based solutions for fall detection and
  prevention: the FARSEEING approach}.
\newblock \bibinfo{journal}{\emph{Zeitschrift f{\"u}r Gerontologie und
  Geriatrie}} \bibinfo{volume}{45}, \bibinfo{number}{8} (\bibinfo{year}{2012}),
  \bibinfo{pages}{722--727}.
\newblock


\bibitem[\protect\citeauthoryear{Mubashir, Shao, and Seed}{Mubashir
  et~al\mbox{.}}{2013}]%
        {mubashir2013survey}
\bibfield{author}{\bibinfo{person}{Muhammad Mubashir}, \bibinfo{person}{Ling
  Shao}, {and} \bibinfo{person}{Luke Seed}.} \bibinfo{year}{2013}\natexlab{}.
\newblock \showarticletitle{A survey on fall detection: Principles and
  approaches}.
\newblock \bibinfo{journal}{\emph{Neurocomputing}}  \bibinfo{volume}{100}
  (\bibinfo{year}{2013}), \bibinfo{pages}{144--152}.
\newblock


\bibitem[\protect\citeauthoryear{Musci, De~Martini, Blago, Facchinetti, and
  Piastra}{Musci et~al\mbox{.}}{2018}]%
        {musci2018online}
\bibfield{author}{\bibinfo{person}{Mirto Musci}, \bibinfo{person}{Daniele
  De~Martini}, \bibinfo{person}{Nicola Blago}, \bibinfo{person}{Tullio
  Facchinetti}, {and} \bibinfo{person}{Marco Piastra}.}
  \bibinfo{year}{2018}\natexlab{}.
\newblock \showarticletitle{Online fall detection using recurrent neural
  networks}.
\newblock \bibinfo{journal}{\emph{arXiv preprint arXiv:1804.04976}}
  (\bibinfo{year}{2018}).
\newblock


\bibitem[\protect\citeauthoryear{Palani, Holt, and Smith}{Palani
  et~al\mbox{.}}{2016}]%
        {palani2016invisible}
\bibfield{author}{\bibinfo{person}{Kartik Palani}, \bibinfo{person}{Emily
  Holt}, {and} \bibinfo{person}{Sean Smith}.} \bibinfo{year}{2016}\natexlab{}.
\newblock \showarticletitle{Invisible and forgotten: Zero-day blooms in the
  IoT}. In \bibinfo{booktitle}{\emph{Pervasive Computing and Communication
  Workshops}}. IEEE.
\newblock


\bibitem[\protect\citeauthoryear{Pedregosa, Varoquaux, Gramfort, Michel,
  Thirion, Grisel, Blondel, Prettenhofer, Weiss, Dubourg,
  et~al\mbox{.}}{Pedregosa et~al\mbox{.}}{2011}]%
        {pedregosa2011scikit}
\bibfield{author}{\bibinfo{person}{Fabian Pedregosa}, \bibinfo{person}{Ga{\"e}l
  Varoquaux}, \bibinfo{person}{Alexandre Gramfort}, \bibinfo{person}{Vincent
  Michel}, \bibinfo{person}{Bertrand Thirion}, \bibinfo{person}{Olivier
  Grisel}, \bibinfo{person}{Mathieu Blondel}, \bibinfo{person}{Peter
  Prettenhofer}, \bibinfo{person}{Ron Weiss}, \bibinfo{person}{Vincent
  Dubourg}, {et~al\mbox{.}}} \bibinfo{year}{2011}\natexlab{}.
\newblock \showarticletitle{Scikit-learn: Machine learning in Python}.
\newblock \bibinfo{journal}{\emph{Journal of Machine Learning Research}}
  \bibinfo{volume}{12} (\bibinfo{year}{2011}), \bibinfo{pages}{2825--2830}.
\newblock


\bibitem[\protect\citeauthoryear{Pratama, Hidayat, et~al\mbox{.}}{Pratama
  et~al\mbox{.}}{2012}]%
        {pratama2012smartphone}
\bibfield{author}{\bibinfo{person}{Azkario~Rizky Pratama},
  \bibinfo{person}{Risanuri Hidayat}, {et~al\mbox{.}}}
  \bibinfo{year}{2012}\natexlab{}.
\newblock \showarticletitle{Smartphone-based pedestrian dead reckoning as an
  indoor positioning system}. In \bibinfo{booktitle}{\emph{International
  Conference on System Engineering and Technology}}. IEEE.
\newblock


\bibitem[\protect\citeauthoryear{Santoyo-Ram{\'o}n, Casilari, and
  Cano-Garc{\'\i}a}{Santoyo-Ram{\'o}n et~al\mbox{.}}{2018}]%
        {santoyo2018analysis}
\bibfield{author}{\bibinfo{person}{Jos{\'e}~Antonio Santoyo-Ram{\'o}n},
  \bibinfo{person}{Eduardo Casilari}, {and} \bibinfo{person}{Jos{\'e}~Manuel
  Cano-Garc{\'\i}a}.} \bibinfo{year}{2018}\natexlab{}.
\newblock \showarticletitle{Analysis of a smartphone-based architecture with
  multiple mobility sensors for fall detection with supervised learning}.
\newblock \bibinfo{journal}{\emph{Sensors}} \bibinfo{volume}{18},
  \bibinfo{number}{4} (\bibinfo{year}{2018}), \bibinfo{pages}{1155}.
\newblock


\bibitem[\protect\citeauthoryear{Schoenmakers}{Schoenmakers}{2018}]%
        {schoenmakers:mpyc}
\bibfield{author}{\bibinfo{person}{Berry Schoenmakers}.}
  \bibinfo{year}{2018}\natexlab{}.
\newblock \showarticletitle{{MPyC}---{Python} package for secure multiparty
  computation}. In \bibinfo{booktitle}{\emph{Workshop on the Theory and
  Practice of MPC}}.
\newblock
\newblock
\shownote{\url{https://github.com/lschoe/mpyc}.}


\bibitem[\protect\citeauthoryear{Shamir}{Shamir}{1979}]%
        {shamir1979share}
\bibfield{author}{\bibinfo{person}{Adi Shamir}.}
  \bibinfo{year}{1979}\natexlab{}.
\newblock \showarticletitle{How to share a secret}.
\newblock \bibinfo{journal}{\emph{Commun. ACM}} \bibinfo{volume}{22},
  \bibinfo{number}{11} (\bibinfo{year}{1979}), \bibinfo{pages}{612--613}.
\newblock


\bibitem[\protect\citeauthoryear{Stikic, Huynh, Van~Laerhoven, and
  Schiele}{Stikic et~al\mbox{.}}{2008}]%
        {stikic2008adl}
\bibfield{author}{\bibinfo{person}{Maja Stikic}, \bibinfo{person}{T{\^a}m
  Huynh}, \bibinfo{person}{Kristof Van~Laerhoven}, {and} \bibinfo{person}{Bernt
  Schiele}.} \bibinfo{year}{2008}\natexlab{}.
\newblock \showarticletitle{ADL recognition based on the combination of RFID
  and accelerometer sensing}. In \bibinfo{booktitle}{\emph{Pervasive Computing
  Technologies for Healthcare}}. IEEE, \bibinfo{pages}{258--263}.
\newblock


\bibitem[\protect\citeauthoryear{Sucerquia, L{\'o}pez, and
  Vargas-Bonilla}{Sucerquia et~al\mbox{.}}{2018}]%
        {sucerquia2018real}
\bibfield{author}{\bibinfo{person}{Angela Sucerquia},
  \bibinfo{person}{Jos{\'e}~David L{\'o}pez}, {and}
  \bibinfo{person}{Jes{\'u}s~Francisco Vargas-Bonilla}.}
  \bibinfo{year}{2018}\natexlab{}.
\newblock \showarticletitle{Real-life/real-time elderly fall detection with a
  triaxial accelerometer}.
\newblock \bibinfo{journal}{\emph{Sensors}} \bibinfo{volume}{18},
  \bibinfo{number}{4} (\bibinfo{year}{2018}), \bibinfo{pages}{1101}.
\newblock


\bibitem[\protect\citeauthoryear{Tao, Kudo, and Nonaka}{Tao
  et~al\mbox{.}}{2012}]%
        {tao2012privacy}
\bibfield{author}{\bibinfo{person}{Shuai Tao}, \bibinfo{person}{Mineichi Kudo},
  {and} \bibinfo{person}{Hidetoshi Nonaka}.} \bibinfo{year}{2012}\natexlab{}.
\newblock \showarticletitle{Privacy-preserved behavior analysis and fall
  detection by an infrared ceiling sensor network}.
\newblock \bibinfo{journal}{\emph{Sensors}} \bibinfo{volume}{12},
  \bibinfo{number}{12} (\bibinfo{year}{2012}), \bibinfo{pages}{16920--16936}.
\newblock


\bibitem[\protect\citeauthoryear{Thang, Viet, Thuc, and Choi}{Thang
  et~al\mbox{.}}{2012}]%
        {thang2012gait}
\bibfield{author}{\bibinfo{person}{Hoang~Minh Thang}, \bibinfo{person}{Vo~Quang
  Viet}, \bibinfo{person}{Nguyen~Dinh Thuc}, {and} \bibinfo{person}{Deokjai
  Choi}.} \bibinfo{year}{2012}\natexlab{}.
\newblock \showarticletitle{Gait identification using accelerometer on mobile
  phone}. In \bibinfo{booktitle}{\emph{Control, Automation and Information
  Sciences}}. IEEE, \bibinfo{pages}{344--348}.
\newblock


\bibitem[\protect\citeauthoryear{Vilarinho, Farshchian, Bajer, Dahl, Egge,
  Hegdal, L{\o}nes, Slettevold, and Weggersen}{Vilarinho et~al\mbox{.}}{2015}]%
        {vilarinho2015combined}
\bibfield{author}{\bibinfo{person}{Thomas Vilarinho}, \bibinfo{person}{Babak
  Farshchian}, \bibinfo{person}{Daniel~Gloppestad Bajer},
  \bibinfo{person}{Ole~Halvor Dahl}, \bibinfo{person}{Iver Egge},
  \bibinfo{person}{Sondre~Steinsland Hegdal}, \bibinfo{person}{Andreas
  L{\o}nes}, \bibinfo{person}{Johan~N Slettevold}, {and}
  \bibinfo{person}{Sam~Mathias Weggersen}.} \bibinfo{year}{2015}\natexlab{}.
\newblock \showarticletitle{A combined smartphone and smartwatch fall detection
  system}. In \bibinfo{booktitle}{\emph{IEEE International Conference on
  Ubiquitous Computing and Communications}}. IEEE, \bibinfo{pages}{1443--1448}.
\newblock


\bibitem[\protect\citeauthoryear{Wang, Liu, Yu, Jia, and Gan}{Wang
  et~al\mbox{.}}{2018}]%
        {wang2018pedestrian}
\bibfield{author}{\bibinfo{person}{Boyuan Wang}, \bibinfo{person}{Xuelin Liu},
  \bibinfo{person}{Baoguo Yu}, \bibinfo{person}{Ruicai Jia}, {and}
  \bibinfo{person}{Xingli Gan}.} \bibinfo{year}{2018}\natexlab{}.
\newblock \showarticletitle{Pedestrian dead reckoning based on motion mode
  recognition using a smartphone}.
\newblock \bibinfo{journal}{\emph{Sensors}} \bibinfo{volume}{18},
  \bibinfo{number}{6} (\bibinfo{year}{2018}), \bibinfo{pages}{1811}.
\newblock


\bibitem[\protect\citeauthoryear{Yavuz, Kocak, Ergun, Alemdar, Yalcin, Incel,
  and Ersoy}{Yavuz et~al\mbox{.}}{2010}]%
        {yavuz2010smartphone}
\bibfield{author}{\bibinfo{person}{Gokhan Yavuz}, \bibinfo{person}{Mustafa
  Kocak}, \bibinfo{person}{Gokberk Ergun}, \bibinfo{person}{Hande~Ozgur
  Alemdar}, \bibinfo{person}{Hulya Yalcin}, \bibinfo{person}{Ozlem~Durmaz
  Incel}, {and} \bibinfo{person}{Cem Ersoy}.} \bibinfo{year}{2010}\natexlab{}.
\newblock \showarticletitle{A smartphone based fall detector with online
  location support}. In \bibinfo{booktitle}{\emph{International Workshop on
  Sensing for App Phones}}. \bibinfo{pages}{31--35}.
\newblock


\bibitem[\protect\citeauthoryear{Zhang, Tian, and Capezuti}{Zhang
  et~al\mbox{.}}{2012}]%
        {zhang2012privacy}
\bibfield{author}{\bibinfo{person}{Chenyang Zhang}, \bibinfo{person}{Yingli
  Tian}, {and} \bibinfo{person}{Elizabeth Capezuti}.}
  \bibinfo{year}{2012}\natexlab{}.
\newblock \showarticletitle{Privacy preserving automatic fall detection for
  elderly using RGBD cameras}. In \bibinfo{booktitle}{\emph{International
  Conference on Computers for Handicapped Persons}}. Springer,
  \bibinfo{pages}{625--633}.
\newblock


\bibitem[\protect\citeauthoryear{Zhang, Cho, Wang, Hsu, Chen, and Shieh}{Zhang
  et~al\mbox{.}}{2014}]%
        {zhang2014iot}
\bibfield{author}{\bibinfo{person}{Zhi-Kai Zhang}, \bibinfo{person}{Michael
  Cheng~Yi Cho}, \bibinfo{person}{Chia-Wei Wang}, \bibinfo{person}{Chia-Wei
  Hsu}, \bibinfo{person}{Chong-Kuan Chen}, {and} \bibinfo{person}{Shiuhpyng
  Shieh}.} \bibinfo{year}{2014}\natexlab{}.
\newblock \showarticletitle{IoT security: ongoing challenges and research
  opportunities}. In \bibinfo{booktitle}{\emph{7th International Conference on
  Service-Oriented Computing and Applications}}. IEEE,
  \bibinfo{pages}{230--234}.
\newblock


\end{thebibliography}





\end{document}